\documentclass[12pt,english]{article} 
\pdfoutput=1
\usepackage[T1]{fontenc}
\usepackage[utf8]{inputenc}
\usepackage{cite}
\usepackage{graphicx} 
\usepackage{dcolumn} 
\usepackage{bm} 
\usepackage{amssymb}
\usepackage{amsmath}
\usepackage{slashed}
\usepackage{mathtools}
\usepackage{hyperref}
\usepackage{cleveref}
\usepackage{geometry}
 \usepackage{relsize}

\usepackage[dvipsnames]{xcolor}

\allowdisplaybreaks

\usepackage{tikz,environ}
\usetikzlibrary{decorations}
\usetikzlibrary{arrows}
\usetikzlibrary{decorations.markings}
\tikzset{
	hard/.style={postaction={decorate},
		line width=0.5mm
	},
		soft/.style={postaction={decorate},
		line width=0.5mm, dashed
	},
	momentum/.style={postaction={decorate},
	line width=0.5mm,
	color=gray,
	decoration={
    markings,
    mark=at position 0.8 with {\arrow{stealth}}}
    },
	hardarrow/.style={postaction={decorate},
	line width=0.5mm,
	decoration={
    markings,
    mark=at position 0.6 with {\arrow{stealth}}}},
}

\newcommand{\Lagr}{\mathcal{L}}

\newcommand{\cM}{\mathcal{M}}

\newcommand{\bS}[1]{{\textbf{#1}}}

\begin{document}

\hypersetup{citecolor=blue,linkcolor=blue,urlcolor=blue}

\interfootnotelinepenalty=10000
\baselineskip=18pt

\hfill CALT-TH-2024-011, CERN-TH-2024-035

\vspace{2cm}
\thispagestyle{empty}
\begin{center}
{\LARGE \bf
Soft Scalars in Effective Field Theory \\
}
\bigskip\vspace{1cm}
\begin{center}{\large Maria Derda,$^{a}$ Andreas Helset,$^{b}$ and Julio Parra-Martinez${}^{c}$}
\end{center}
\vspace{7mm}
{
\it ${}^a$Walter Burke Institute for Theoretical Physics,\\[-1mm]
California Institute of Technology, Pasadena, California 91125\\[1.5mm]
${}^b$Theoretical Physics Department, CERN, 1211 Geneva 23, Switzerland\\[1.5mm]
${}^c$Department of Physics and Astronomy,\\[-1mm]  University of British Columbia, Vancouver, V6T 1Z1, Canada}
\end{center}

\bigskip\vspace{1cm}
\centerline{\large\bf Abstract}
\begin{quote} \small

We derive a soft theorem for a massless scalar in an effective field theory with generic field content using the geometry of field space. This result extends the geometric soft theorem for scalar effective field theories by allowing the massless scalar to couple to other scalars, fermions, and gauge bosons. The soft theorem keeps its geometric form, but where the field-space geometry now involves the full field content of the theory. As a bonus, we also present novel double soft theorems with fermions, which mimic the geometric structure of the double soft theorem for scalars.

\end{quote}

\setcounter{footnote}{0}

\newpage

\tableofcontents
\newpage

\section{\label{sec:intro}Introduction}

Low-energy modes are often related to the symmetry properties of a theory. In scattering amplitudes, this connection takes the form of a soft limit, where the momentum of a particle is sent to zero. If this limit exhibits a universal pattern, we declare it a soft theorem. Salient examples of such relations are the pion soft theorem---the Adler zero~\cite{Adler:1964um}---which is a consequence of the spontaneously-broken chiral symmetry, the soft theorem for gauge theories~\cite{Low:1958sn,Weinberg:1965nx,Burnett:1967km}, which follows from charge conservation, and the graviton soft theorem~\cite{Weinberg:1965nx}, due to energy-momentum conservation. In general, a theory with a nonlinearly realized symmetry manifests this fact in scattering amplitudes through soft theorems. Also in condensed matter systems, such as solids, fluids, and superfluids, phonon soft theorems are direct consequences of spontaneous symmetry breaking~\cite{Green:2022slj,Cheung:2023qwn}. Finally, there is a close connection between soft theorems and asymptotic symmetries~\cite{Strominger:2017zoo,Strominger:2013jfa,Kapec:2015vwa,Campiglia:2015qka,Himwich:2019dug,He:2019jjk,Kapec:2021eug,Kapec:2022axw,Kapec:2022hih,Adamo:2023zeh,Chen:2023tvj} (see also ref.~\cite{Pasterski:2021raf} and references therein).

However, symmetry is not the only possible origin of these universal relations between scattering amplitudes. A geometric soft theorem for scalar effective field theories was derived solely as a consequence of the geometry of field space~\cite{Cheung:2021yog}, which did not rely on any symmetry of the theory. In the simplest case with no potential, the geometric soft theorem takes the form
\begin{align}
    \label{eq:geometric_soft_theorem}
    \lim_{q\rightarrow 0} A_{n+1} = \nabla_{i} A_{n} ,
\end{align}
where $A_{n}$ is an $n$-particle scattering amplitude, $\nabla_{i}$ is the field-space covariant derivative with respect to the vacuum expectation value (VEV), and $i$ is the flavor index of the soft scalar. Mathematically,  as explained in ref.~\cite{Cheung:2021yog}, scattering amplitudes of scalars take values in the tangent bundle of the field-space manifold and the soft theorem is described by the familiar Levi-Civita connection on the tangent space. This geometric picture is general for any effective field theory and manifests the invariance of scattering amplitudes under changes of field basis.\footnote{The geometric soft theorem has also found an interpretation in the context of celestial holography~\cite{Kapec:2022axw,Kapec:2022hih}.}

In this paper, we extend the geometric soft theorem for a massless scalar by allowing the scalar to couple to fermions and gauge bosons. The geometry must be extended to include the full field content of the theory, since we can perform field redefinitions for any field in our theory. Remarkably, this is precisely what we need to complete the geometric soft theorem, which takes a form similar to \cref{eq:geometric_soft_theorem} but with the upgrade 
\begin{equation}
    \nabla \rightarrow \bar \nabla = \partial + \Gamma^s + \Gamma^f + \Gamma^g\,,
\end{equation} i.e., the covariant derivative for the full field-space geometry which includes a connection for scalars, $\Gamma^s$, fermions, $\Gamma^f$, and gauge bosons, $\Gamma^g$. More precisely, the additional fields take values in a vector bundle over the field space, with an associated connection which features in the soft theorem.

We can also reverse this logic and use the new geometric soft theorem as justification for the extension of the geometric picture to include particles with spin. For example, the scalar soft theorem for a theory of scalars and fermions involves the connection $\bar \Gamma^{p}_{i r}$, where $i$ is a scalar flavor index and $p,r$ are fermion flavor indices. This shows that the definition of a scalar-fermion geometry is not simply a formal exercise but that it has physical consequences manifested in the soft scalar limit.

The geometric soft theorems have wide applicability and are realized in many theories of interest. For instance, when the massless scalars are Nambu-Goldstone bosons (NGBs), they generalize the Adler zero and describe the coupling of NGBs to other species. They also describe the dependence of amplitudes in supersymmetric theories on the VEV of scalar moduli \cite{Wang:2015jna, Wang:2015aua, Lin:2015dsa}.
Furthermore, they provide a vast generalization of the well-known low-energy theorems for a light Higgs (see, e.g., ref.~\cite{Kniehl:1995tn}). This is, of course, not an exhaustive list.

The paper is organized as follows. First, we review the geometry of field space for scalars, fermions, and gauge bosons. Then we derive the geometric soft theorem, valid for any effective field theory with a massless scalar. We present the geometric soft theorem in three parts, first with only scalars, then with fermions, and last with gauge bosons. In the following section, we present a novel double soft theorem, where the momenta of two scalars are sent to zero. In this case, the soft theorem involves the curvature of the full field-space geometry, including components for fermions and gauge bosons. Then we present new double-soft theorems for fermions. These soft theorems are almost identical to the double-scalar soft theorem, up to the simple replacement of a kinematic factor. 
Numerous examples are listed in \cref{sec:examples}. 
We end with a discussion and outlook.



\section{\label{sec:field-space}Geometry of Field Space}

We consider an effective theory that includes scalars, fermions, and gauge bosons. To low orders in the derivative expansion, the Lagrangian is
\begin{align}
    \label{eq:general_Lagrangian}
    \Lagr =& \frac{1}{2} h_{IJ}(\phi) (D_{\mu} \phi)^{I} (D^{\mu} \phi)^{J} - V(\phi) 
    + i \frac{1}{2} k_{\bar pr}(\phi) (\bar \psi^{\bar p} \gamma^{\mu}\overset{\leftrightarrow}{D}_{\mu} \psi^{r}) \nonumber \\
    &+ i \omega_{\bar p r I}(\phi) (\bar \psi^{\bar p} \gamma_{\mu} \psi^{r}) (D^{\mu} \phi)^{I} - \bar \psi^{\bar p} \cM(\phi)_{\bar pr}  \psi^{r} + c_{\bar p r \bar s t}(\phi) (\bar \psi^{\bar p} \gamma_{\mu}  \psi^{r}) (\bar \psi^{\bar s} \gamma^{\mu}  \psi^{t})  \nonumber \\
    &- \frac{1}{4} g_{AB}(\phi) F_{\mu\nu}^{A} F^{B\mu\nu} + d_{\bar p r A}(\phi) (\bar \psi^{\bar p} \sigma^{\mu\nu} \psi^{r})F^{A}_{\mu\nu} + \dots,
\end{align}
where we allow for higher-derivative operators and operators with more fermion fields, but do not list them explicitly. To keep the discussion simple, we omit the CP-odd scalar--gauge-boson couplings, $-\tfrac{1}{4}\tilde g_{AB}(\phi) F_{\mu\nu}^{A} \tilde F^{B\mu\nu}$, but all results generalize straightforwardly when they are present.  
We group all fields with the same spin into multiplets, with scalar indices $I,J,\dots$, fermion indices $p,\bar p,\dots$, and gauge indices $A,B,\dots$
The functions $h_{IJ}(\phi)$, $V(\phi)$, $k_{\bar pr}(\phi)$ etc., are functions of the scalar fields. By introducing these functions, we group infinite families of higher-dimensional operators into compact structures \cite{Helset:2020yio}. This grouping of operators underlies the geometric construction. The scalars $\phi^{I}$ and fermions $\psi^{r}$ can be charged under the gauge symmetry through the covariant derivative, which we describe in more detail below.

As in any effective field theory, the number of independent operators is less than the number of possible composite operators consistent with the symmetries of the theory. This is because integration-by-parts relations and field redefinitions can be used to write the Lagrangian in a form with a minimal number of operators, i.e., a nonredundant operator basis. This freedom of redefining the fields (at least when the field redefinition does not involve derivatives) takes on a geometric meaning, paralleling coordinate changes in differential geometry. The field-space geometry for scalars is by now a standard quantum-field-theory technique. See refs.~\cite{Meetz:1969as,Honerkamp:1971sh,Ecker:1971xko,osti_4340109,Tataru:1975ys} for some early works on the connection between differential geometry and field redefinitions and refs.~\cite{Alonso:2015fsp,Alonso:2016oah,Alonso:2017tdy,Helset:2018fgq,Finn:2019aip,Helset:2020yio,Cohen:2020xca,Cohen:2021ucp,Helset:2022pde,Alonso:2022ffe,Alonso:2023upf,Jenkins:2023bls} for modern applications of the scalar geometry in effective field theories. Recently, this geometric picture has been extended to include both fermions and gauge bosons \cite{Finn:2020nvn,Cheung:2022vnd,Helset:2022tlf,Assi:2023zid}, and several proposals attempt to extend the geometric description to accommodate field redefinitions with derivatives \cite{Cheung:2022vnd,Cohen:2022uuw,Craig:2023wni,Craig:2023hhp,Alminawi:2023qtf,Cohen:2023ekv}.

\subsection{Scalars}

The geometry of the scalar field space is dictated by the metric $h_{IJ}$. From this metric, we can derive the Christoffel symbol
\begin{align}
    \label{eq:scalar_Christoffel}
    \Gamma^{I}_{JK} = \frac{1}{2} h^{IL} \left( h_{JL,K} + h_{LK,J} - h_{JK,L} \right),
\end{align}
where $h_{IJ,K} = \partial_{K} h_{IJ}$, and the Riemann curvature
\begin{align}
    R_{IJKL} = h_{IM} \left( \partial_{K}\Gamma^{M}_{LJ} + \Gamma^{M}_{KN} \Gamma^{N}_{LJ} - (K \leftrightarrow L) \right) .
\end{align}
The covariant derivative $\nabla_{I}$ uses the connection in \cref{eq:scalar_Christoffel}. 
The field-space geometry for scalars captures field redefinitions of the form $\phi \rightarrow F(\phi)$, where $F'(v)\neq 0$ at the VEV $v^{I}$, and was used to describe the geometric soft theorem for scalar effective field theories~\cite{Cheung:2021yog}. 

The scalar field in the Lagrangian, $\phi^{I}$, can be used as an interpolating field between the vacuum and a one-particle state,
\begin{align}
    \langle p_{i}| \phi^{I}(x)|0\rangle = e^{I}_{i}(v) e^{ip\cdot x} ,
\end{align}
where the momentum is on the mass shell, $p^2 = m_{i}^{2}(v)$, and $e^{I}_{i}(v)$ is the tetrad, which is defined from the metric
\begin{align}
    h_{IJ}(v) = e_{Ii}(v) e^{i}_{J}(v) .
\end{align}
The tetrad is the wavefunction factor in the LSZ reduction formula. Its role is to canonically normalize and rotate between the flavor-eigenstate fields in the Lagrangian and the mass eigenstates used in scattering amplitudes. Therefore, a scattering amplitude is a tensor with lowercase tetrad indices.
Further details on the geometric construction for scalars can be found in ref.~\cite{Cheung:2021yog}.

\subsection{Fermions}

We follow the setup in ref.~\cite{Assi:2023zid} to describe fermions geometrically. A similar approach, but with certain differences in the technical steps, is described in refs.~\cite{Finn:2020nvn,Gattus:2023gep}.
The main novelty for the fermion geometry compared with the scalar geometry discussed above is that we now must accommodate anticommuting fields into the geometric picture. This can be conveniently done by replacing the Riemannian manifold with a supermanifold, which involves Grassmann coordinates \cite{DeWitt:2012mdz}. Note that the notion of a supermanifold is distinct from supersymmetry, and we do not require our theories to possess supersymmetry. 

The fermion geometry is defined by the metric\footnote{Following the conventions of ref.~\cite{DeWitt:2012mdz}, the metric should be written as $\,_{\mathfrak{i}} \bar g_{\mathfrak{j}}$, and shifting the indices to the right will pick up additional signs, $\bar g_{\mathfrak{ij}} = (-1)^{\mathfrak{i}}\;_{\mathfrak{i}}\bar g_{\mathfrak{j}}$. Here, we exclusively deal with the metric and allow ourselves to abuse the notation by having the indices on the right from the start.}
\begin{align}
    \label{eq:fermion_metric}
    \bar g_{\mathfrak{ij}} = \begin{pmatrix}
    h_{IJ} & (\bar \psi \omega^{-})_{rI} &  (\omega^{+}\psi)_{\bar r I} \\
    - (\bar \psi \omega^{-})_{pJ}  & 0 & k_{\bar r p} + c_{\bar r p} \\
    - (\omega^{+} \psi)_{\bar p J} & - k_{\bar p r} - c_{\bar p r} & 0
    \end{pmatrix},
\end{align}
where $\omega^{\pm}_{\bar p r I} = \omega_{\bar p r I} \pm \tfrac{1}{2} k_{\bar p r, I}$. The scalar indices $I,J,\dots$ and the fermion indices $p,\bar p,\dots$ are unified in the indices $\mathfrak{i},\mathfrak{j},\dots$. The metric and descendant quantities are denoted with a bar to distinguish them from the corresponding quantities in the scalar geometry.

Four-fermion operators were not included in the geometric construction in ref.~\cite{Assi:2023zid}. We include them in the metric in \cref{eq:fermion_metric} through the term $c_{\bar p r} = 4(c_{\bar p r \bar s t} + c_{\bar p t \bar s r})\psi^{t} \bar \psi^{\bar s}$. There are several reasons why this construction is sensible. First, the four-fermion operators transform as tensors under redefinitions of the fermion fields that depend on the scalar fields. Thus, they are fine objects to add to the metric, as they do not spoil any of the transformation properties used to bootstrap the metric for the two-fermion sector. Second, the other operators which make up the scalar-fermion metric are combinations of two scalar currents or one scalar current and one fermion current. Thus, it is natural to expect that operators with two fermion currents can also reside in the metric. Lastly, in the supersymmetric nonlinear sigma model, the coefficient of the four-fermion operator is the Riemann curvature. Therefore, these operators must be included in the metric even for a general theory without supersymmetry, since the supersymmetric theory should be obtainable from the general theory by picking the correct field content and tuning the coefficients. The virtue of this definition will be apparent when we consider single and double soft theorems of scalars and fermions.

From this metric we can also calculate the Christoffel symbol and the curvature, but with the definitions appropriate for a supermanifold. In particular, the relevant connection coefficients are \cite{Assi:2023zid} 
\begin{align}
    \label{eq:fermion_Christoffel0}
    \bar \Gamma^{K}_{IJ} &= \Gamma^{K}_{IJ},  \\
    \label{eq:fermion_Christoffel1}
    \bar \Gamma^{p}_{Ir} &= k^{p\bar s} \omega^{+}_{\bar s r I},  \\
    \label{eq:fermion_Christoffel2}
    \bar \Gamma^{\bar p}_{I\bar r} &= - \omega^{-}_{\bar r s I} k^{s \bar p}, 
\end{align}
and the corresponding curvatures are 
\begin{align}
    \label{eq:scalar-fermion_curvature}
    \bar R_{\bar p r IJ} &= \omega^{}_{\bar p r J, I} + \omega^{-}_{\bar p s I} k^{s \bar t} \omega^{+}_{\bar t r J} - (I \leftrightarrow J) , \\
    \label{eq:fermion-fermion_curvature}
    \bar R_{\bar p r \bar s t} &= 4 (c_{\bar p r \bar s t} + c_{\bar p t \bar s r}) ,
\end{align}
all evaluated at the VEV. The covariant derivative $\bar \nabla$ uses the connections in \cref{eq:fermion_Christoffel0,eq:fermion_Christoffel1,eq:fermion_Christoffel2}. For our purposes, where we analyze the geometric structure of scattering amplitudes, we only need the geometric quantities evaluated at the VEV. Other applications, such as background-field calculations~\cite{Helset:2022pde,Assi:2023zid,Jenkins:2023bls}, also use the geometric information away from the VEV.

Similar to the scalars above, the flavor-basis field $\bar \psi^{\bar R}$ sandwiched between the one-particle fermion state and the vacuum is
\begin{align}
    \langle p_{\bar r}| \bar\psi^{\bar R}(x) | 0 \rangle = \bar u(p) e^{\bar R}_{\bar r}(v) e^{i p\cdot x}.
\end{align}
Note that we here used capital indices for the flavor-basis field $\bar \psi^{\bar R}$ to distinguish them from the lowercase indices mass-eigenstate basis. However, for esthetic reasons, we used lowercase indices in the Lagrangian in \cref{eq:general_Lagrangian}. Hopefully, this slight abuse of notation will not cause confusion. The tetrad, which is derived from the metric, will implicitly be used to transform between the two bases,
\begin{align}
    \begin{pmatrix}
        0 & k_{R \bar P} \\
        - k_{\bar P R} & 0
    \end{pmatrix} 
    = e_{\bar P}^{\bar p} 
    \begin{pmatrix}
        0 & \delta_{r \bar p} \\
        - \delta_{\bar p r} & 0
    \end{pmatrix}
    e_{R}^{r} ,
\end{align}
where $\delta_{\bar p r}$ is the Kronecker delta.
The fermions are canonically normalized and rotated to the mass-eigenstate basis via the tetrad. The tetrad shows up in the LSZ reduction formula for the fermions as the wavefunction factor, exactly as for the scalars.

\subsection{Gauge bosons}

There is a larger freedom in how to construct a geometric field space which includes gauge bosons. One option is to use the geometry-kinematics map \cite{Cheung:2022vnd}, where essentially the gauge bosons act like scalars, and all the geometric quantities in the scalar field space get upgraded to depend on both the scalars and the gauge fields. As an added bonus, the geometry-kinematics duality allows all higher-derivative operators to be placed on the same footing as the two-derivative operators, thus providing a geometric understanding of derivative field redefinitions. 
The advantage of using the geometry-kinematics map is that statements which hold for scalar effective field theories immediately get upgraded to statements which hold for general bosonic effective field theory. This includes the geometric soft theorem. Some drawbacks of this approach are that the notation is rather compact and that there are some ambiguities in the initial choice for the metric. 

Another option is to treat the gauge fields separately from the scalar fields. One such formulation was introduced in ref.~\cite{Helset:2022tlf}. By using a geometric gauge fixing \cite{Helset:2018fgq}, the metric takes the form
\begin{align}
    \label{eq:gauge_metric}
    \bar g_{\alpha\beta} = \begin{pmatrix}
        \bar g_{\mathfrak{ij}} & 0 \\
        0 & g_{AB} 
    \end{pmatrix} ,
\end{align}
where we also include the fermions via the metric in \cref{eq:fermion_metric}. 
For a theory without fermions, we simply replace $\bar g_{\mathfrak{ij}}\rightarrow h_{IJ}$ in \cref{eq:gauge_metric}. 
We have stripped off a factor $(-\eta_{\mu\nu})$ compared to the metric in ref.~\cite{Helset:2022tlf}. This factor can be trivially reinstated with the replacement $g_{AB}\to -g_{AB} \eta_{\mu_A\mu_B}$. 
The indices $\alpha,\beta,\dots$ include all scalar and fermion indices, as well as the gauge-field indices $A,B,\dots$ Here we slightly abuse the notation by denoting the full scalar--fermion--gauge-boson metric with a bar, as we did in the scalar-fermion metric in \cref{eq:fermion_metric}. 
If we included the CP-odd scalar--gauge-boson couplings, $-\tfrac{1}{4}\tilde g_{AB}(\phi) F_{\mu\nu}^{A} \tilde F^{B\mu\nu}$, the metric in \cref{eq:gauge_metric} would change to $g_{AB} \rightarrow g_{AB}^{\pm} = g_{AB} \pm \tilde g_{AB}$ for positive/negative helicity gauge fields. This is analogous to how positive/negative helicity fermions couple through the vertex $\omega^{\pm}_{\bar pr I}$. For simplicity, we omit the CP-odd couplings in the gauge metric.

In this paper, we opt for using \cref{eq:gauge_metric} for concreteness. This cleanly separates the particles of different spin. However, we will in passing mention how our results change when using the geometry-kinematics map. Intriguingly, both definitions of the gauge-boson metric lead to a new geometric soft theorem. These soft theorems are equivalent but differ in form.

With this choice of metric, we can calculate the connection \cite{Helset:2022tlf},
\begin{align}
    \label{eq:gauge_Christoffel}
    \bar \Gamma^{a}_{bi} = \frac{1}{2} g^{ac} (\nabla_{i} g_{cb}) ,
\end{align}
and the curvature,
\begin{align}
    \bar R_{aijb} = \frac{1}{2} \nabla_{i} \nabla_{j} g_{ab} - \frac{1}{4} (\nabla_{j} g_{ac}) g^{cd} (\nabla_{i} g_{db}) .
\end{align}

Next, we need to relate the gauge field to the scattering state. The gauge field creates a one-particle state,
\begin{align}
    \langle p_b, \epsilon | A^{B}_{\mu}(x) | 0 \rangle = \epsilon^{*}_{\mu}(p) e^{B}_{b}(v) e^{i p\cdot x} .
\end{align}
The polarization vector $\epsilon_{\mu}$ encodes the two degrees of freedom for a massless gauge field, or the three degrees of freedom for a massive gauge field. Sometimes, we combine the polarization vector and the tetrad, which is defined as $g_{AB}(v) = e_{Aa}(v) e^{a}_{B}(v)$, into a new polarization tensor, $\epsilon^{B}_{b\mu}(v) = \epsilon_{\mu} e^{B}_{b}(v)$, which carry the tetrad indices. The scattering amplitude is multilinear in these polarization vectors, and it will be a tensor with gauge-boson indices in the mass-eigenstate basis.

We also consider massive gauge bosons which get their mass through the Higgs mechanism. As is well known (and reviewed in ref.~\cite{Cheung:2021yog}), a global symmetry in the scalar sector is associated with a set of Killing vectors, $t^I{}_A(\phi)$, such that 
\begin{equation}
    \phi^I \to \phi^I + c^A\, t^I{}_A(\phi)
\end{equation}
leaves the Lagrangian invariant for any $c^A$. The Killing vectors satisfy commutation relations
\begin{equation}
    t^J{}_A(\phi)\partial_J t^I{}_B(\phi) -  t^J{}_B(\phi)\partial_J t^I{}_A(\phi) = f_{AB}{}^C t^I{}_C(\phi)
\end{equation}
corresponding to a Lie algebra\footnote{Note that we use a convention different from the usual one in the amplitudes literature ($[t^A,t^B] = \sqrt{2} f_{AB}{}^C t_C$, see, e.g., ref.~\cite{Dixon:1996wi}), so our gauge-boson amplitudes carry additional factors of $\sqrt{2}$ in comparison.}. When this symmetry is gauged, the covariant derivative which describes the coupling of the scalars to gauge bosons is
\begin{align}
    (D_\mu \phi)^{I} = \partial_{\mu} \phi^{I} + t^{I}_{B}(\phi) A^{B}_{\mu} ,
\end{align}
where $t^{I}_{A}(\phi)$ is a Killing vector of the scalar field-space manifold. The gauge bosons can acquire mass through the Higgs mechanism. Some of the scalar fields then take on a nonvanishing VEV, which spontaneously breaks the gauge symmetry. In this case, some of the Killing vectors are nonzero at the VEV, $t^{I}_{A}(v) \neq 0$. 
The mass of a gauge boson is generally given by the square of the Killing vectors evaluated at the VEV,
\begin{align}
    m_{ab}^{2}(v) = t^I{}_{a}(v) t_{Ib}(v)\,.
\end{align}
However, if the gauge group is not broken, then the Killing vectors vanish at the VEV, $t^{I}_{A}(v) = 0$, and the gauge bosons remain massless. We will not commit to either case, and allow for having both charged and neutral scalars as well as massless and massive gauge bosons in our effective field theory. 

For later reference, it is useful to quote the Goldstone boson equivalence theorem in the geometric notation:
\begin{equation}\label{eq:NBGequiv}
    \lim_{p_1 \to \infty}   A_{a_1 \cdots}(1_L \cdots) = \frac{t^{i_1}{}_{a_1}(v)}{m_a} A_{i_1 \cdots}(1_{\rm scalar} \cdots)
\end{equation}
where the left-hand side is the amplitude of a longitudinal massive gauge boson, and the right-hand side is the amplitude of the ``would-be'' NGB scalar which is eaten in the Higgs mechanism.



\section{\label{sec:soft-theorem}Geometric Soft Theorem}

Below, we present the geometric soft theorem for a massless scalar in a general effective field theory with other (possibly massive) scalars, fermions, and gauge bosons. The derivation of this result is analogous to the derivation for scalar effective field theories \cite{Cheung:2021yog}. We first review the case for scalars before also including fermions and gauge bosons. The general soft theorem is the union of these results.

\subsection{Scalars}

The geometric soft theorem for scalars was derived in ref.~\cite{Cheung:2021yog}. We reproduce it here. It involves the covariant derivative in field space acting on either the lower-point amplitude or the mass matrix of the external particles. The index $j$ on the covariant derivative corresponds to the index of the particle with momentum $q$, whose momentum is sent to zero. In full, the geometric soft theorem is
\begin{align}
    \label{eq:soft_theorem_scalar}
    \lim_{q\rightarrow 0} A_{n+1,i_1\cdots i_{n} j} =& \nabla_{j} A_{n,i_1 \cdots i_n} + \sum^{n}_{a=1} \frac{\nabla_{j} V_{i_a}^{\;\; j_a}}{(p_a+q)^2 - m^2_{j_a}} \left(1 + q^{\mu} \frac{\partial}{\partial p_a^{\mu}} \right) A_{n,i_1 \cdots j_a \cdots i_n},
\end{align}
where $V_{ij} \equiv V_{;ij}$. 
The first term in the soft theorem acts on all coupling constants and masses in the amplitude, which are viewed as functions of the VEV. The second term is essential to be consistent with the on-shell conditions for all particles. 

This geometric soft theorem unifies the Adler zero for Nambu-Goldstone bosons on a symmetric coset \cite{Adler:1964um}, soft theorems for more general Nambu-Goldstone bosons \cite{Kampf:2019mcd}, and the dilaton soft theorem \cite{Callan:1970yg,Boels:2015pta,Huang:2015sla,DiVecchia:2015jaq}. 
For illustration, we have listed in \cref{sec:examples} examples of scattering amplitudes for four and five scalar particles and shown how they are connected through the geometric soft theorem.

\subsection{Fermions}

Next, we add fermions to the mix. The geometric soft theorem for a massless scalar in the presence of both scalars and fermions is new. It bears stark resemblance to the soft theorem above. The geometric soft theorem again depends on the covariant derivative in field space, but this time for the combined scalar-fermion geometry defined through the metric in \cref{eq:fermion_metric}. This covariant derivative $\bar \nabla_{i}$ is denoted with a bar to indicate that it is also sensitive to fermionic flavor indices.

The full scalar-fermion soft theorem is
\begin{align}
    \label{eq:soft_theorem_fermion}
    \lim_{q\rightarrow 0} A_{n+1,\mathfrak{i}_1 \cdots \mathfrak{i}_{n} j} =& \bar \nabla_{j} A_{n,\mathfrak{i}_1 \cdots \mathfrak{i}_n} + \sum_{a\in \{\textrm{scalar}\}} \frac{\bar \nabla_{j} V_{i_a}^{\;\; j_a}}{(p_a+q)^2 - m^2_{j_a}} \left(1 + q^{\mu} \frac{\partial}{\partial p_a^{\mu}} \right) A_{n,\mathfrak{i}_1 \cdots j_a \cdots \mathfrak{i}_n} \nonumber \\
    & + \sum_{b\in \{\textrm{fermion}\}} \sum_{\rm{spin}} \frac{\bar \nabla_{j} \cM^{r_b}_{\quad p_b} (\bar u(p+q) u(p)) }{(p_b+q)^2 - m^2_{r_b}} \left(1 + q^{\mu} \frac{\partial}{\partial p_b^{\mu}} \right) A_{n,\mathfrak{i}_1 \cdots r_b \cdots \mathfrak{i}_n}
    \nonumber \\
    & + \sum_{b\in \{\textrm{anti-fermion}\}} \sum_{\rm{spin}} \frac{\bar \nabla_{j} \cM_{\bar p_b}^{\quad \bar r_b} (-\bar v(p) v(p+q))}{(p_b+q)^2 - m^2_{r_b}} \left(1 + q^{\mu} \frac{\partial}{\partial p_b^{\mu}} \right) A_{n,\mathfrak{i}_1 \cdots \bar r_b \cdots \mathfrak{i}_n} \,,
\end{align}
where $\cM$ is the fermion mass matrix. 
Let us unpack this soft theorem. 
We take all momenta to be incoming and write the amplitude with lowered flavor indices. Note that the tetrads are implicitly included for both scalars and fermions, although we use the same index for the fermion flavors in the amplitude as in the Lagrangian. The tetrads canonically normalize and rotate the states to the mass-eigenstate basis, where the mass matrix is diagonal. 
Another thing we have kept implicit is the label for the spin component of the fermion wave functions. 
The spin is summed over for the external fermion wavefunction in the $n$-point amplitude and the shifted spinor in the prefactors. 

The first line in \cref{eq:soft_theorem_fermion} is similar to the scalar soft theorem in \cref{eq:soft_theorem_scalar}, with the replacement $\nabla_{i} \rightarrow \bar \nabla_{i}$, while the second and third lines are the covariant derivative acting on the external fermion propagators. The last three terms can be unified to the covariant derivative of a single mass matrix, where the indices run over both scalar and fermion flavors, but we choose to write out all the terms explicitly for clarity.

The geometric soft theorem in \cref{eq:soft_theorem_fermion} holds at tree level. However, in the case where the potential and fermion mass matrix vanish, $V(\phi)$=0 and $\cM(\phi)=0$, we believe the soft theorem holds at all loop orders, for the same reasons as in the soft scalar theorem \cite{Cheung:2021yog}.

For this soft theorem to have a sensible on-shell interpretation, it must commute with the on-shell conditions. Consider the action of the soft theorem on the on-shell condition for an incoming fermion,
\begin{align}
    (\slashed p_s \delta^{r}_{\;\; s}- \cM^{r}_{\;\; s} ) u(p_s) = 0 \,.
\end{align}
The covariant derivative shifts the mass matrix,  
\begin{align}
    \label{eq:fermion_onshell1}
    \bar \nabla_{j} (\slashed p_s \delta^{r}_{\;\; s}- \cM^{r}_{\;\; s} ) u(p_s) = - (\bar \nabla_{j} \cM^{r}_{\;\; s} ) u(p_s) ,
\end{align}
and the third term in \cref{eq:soft_theorem_fermion} acts on the spinor,
\begin{align}
    \label{eq:fermion_onshell2}
    & \sum_{\rm{spin}} \frac{\bar \nabla_{j} \cM^{t}_{\;\; s} (\bar u(p_s+q) u(p_s)) }{(p_s+q)^2 - m^2_{r}} \left(1 + q^{\mu} \frac{\partial}{\partial p_s^{\mu}} \right) (\slashed p \delta^{r}_{\;\; t}- \cM^{r}_{\;\; t} ) u(p_s)
    \nonumber \\ 
    =& \sum_{\rm{spin}} \frac{\bar \nabla_{j} \cM^{t}_{\;\; s} (\bar u(p_s+q) u(p_s)) }{(p_s+q)^2 - m^2_{r}} ((\slashed p_s + \slashed q) \delta^{r}_{\;\; t}- \cM^{r}_{\;\; t} ) u(p_s+q)
    = (\bar \nabla_{j} \cM^{r}_{\;\; s})  u(p_s) ,
\end{align}
where we have used that the sum over spins is
\begin{align}
    \sum_{\rm{spin}} u(p) \bar u(p) =  (\slashed p + m) .
\end{align}
Clearly, \cref{eq:fermion_onshell1,eq:fermion_onshell2} cancel, which means that the soft theorem does not spoil the on-shell conditions for incoming fermions and can therefore be applied unambiguously to scattering amplitudes. 

The soft theorem also commutes with the on-shell condition for incoming anti-fermions. In this case, the cancellation happens between the covariant derivative and the fourth term in \cref{eq:soft_theorem_fermion}, where we now have to use that the sum over spins is
\begin{align}
    \sum_{\rm{spin}} v(p) \bar v(p) = (\slashed p - m) .
\end{align}

Perhaps the most well-known case of low-energy dynamics for relativistic scalars is the theory of pions. 
The soft limit of pion--pion scattering vanishes, known as the Adler zero \cite{Adler:1964um}. In contrast, the soft limit of a pion scattering off nucleons does not vanish. However, this limit is universal and can be derived using current algebra methods. The nonzero soft limit of pion-nucleon scattering is related to the coupling in the so-called gradient-coupling theory \cite{Adler:1965ga}. This is nothing but the couplings $\omega^{\pm}$, which enter the geometric soft theorem through the covariant derivative $\bar \nabla$. Thus, the pion-nucleon soft theorem is a special case of the geometric soft theorem \cite{Adler:1965ga,Nambu:1962lbq,Nambu:1962ilv,coleman_1985}. 
Another special case of the geometric soft theorem is the low-energy limit of the $\eta'$ particle in large-$N$ QCD described long ago by Witten \cite{Witten:1979vv}. This is a pseudoscalar NGB for the axial $U(1)$ symmetry of QCD which remains unbroken in the planar limit. Its soft limit computes derivatives of scattering amplitudes with respect to the QCD $\theta$ angle, or equivalently the $\eta'$ VEV.
More generally, the leading term in the soft limit goes as $1/(p\cdot q)$ and comes from the covariant derivative acting on the scalar potential $V(\phi)$ or the fermion mass matrix $\cM(\phi)$. This universal soft behavior of scalars is analogous to the leading soft limit of photons and gravitons, and it follows from similar polology considerations \cite{Campiglia:2017dpg,Biswas:2022lsj}.

The derivation of the geometric soft theorem in \cref{eq:soft_theorem_fermion} is analogous to the derivation for a scalar effective field theory in ref.~\cite{Cheung:2021yog}. We will here highlight the main novelties compared to the scalar case. The derivation begins by using the Euler-Lagrange equations,
\begin{align}
    \partial_{\mu} \mathcal{J}^{\mu}_{I} = \partial_{I} L ,
\end{align}
where
\begin{align}
    \mathcal{J}^{\mu}_{I} = \frac{\delta L}{\delta (\partial_{\mu}\phi^{I})}
    \qquad \textrm{and}
    \qquad
    \partial_{I} L = \frac{\delta L}{\delta \phi^{I}} .
\end{align}
Since the scalar field $\phi^{I}$ is expanded around the VEV $v^{I}$, their appearance in the Lagrangian is identical, and we can equivalently calculate the variation of the Lagrangian with respect to the VEV to obtain $\partial_{I} L$. 

The only terms that will affect the soft theorem are operators that are at most cubic in the field. We split the contributions coming from scalar and fermion operators. We find that
\begin{align}
    \mathcal{J}_{I}^{\mu} =& \left(\mathcal{J}_{I}^{\mu}\right)_{\rm{scalar}} + i \omega_{\bar p r I}(v) (\bar \psi^{\bar p} \gamma^{\mu} \psi^{r}) + \dots ,\\
    \partial_{I}L =& \left(\partial_{I}L\right)_{\rm{scalar}} + \frac{1}{2} i k_{\bar p r, I}(v) (\bar \psi^{\bar p} \overset{\leftrightarrow}{\slashed \partial} \psi^{r}) - \cM_{\bar p r,I}(v) (\bar \psi^{\bar p} \psi^{r}) + \dots
\end{align}
We now collect the contributions from the fermion operators, and insert them on external fermion lines,
\begin{align}
    \label{eq:fermion_operator_insertion}
    \langle \partial_{I}L - \partial_{\mu} \mathcal{J}^{\mu}_{I}\rangle_{\rm{ext.\;fermion}} =& -\langle (\bar \psi \omega^{-})_{tI} (i \delta^{t}_{\;\;r}\slashed \partial - \cM^{t}_{\;\; r})\psi^{r} \rangle_{\rm{ext}}
    -\langle \bar\psi^{\bar p} (i \delta^{\;\;\bar t}_{\bar p}\overset{\leftarrow}{\slashed \partial} + \cM^{\;\;\bar t}_{\bar p}) (\omega^{+}\psi)_{\bar tI}  \rangle_{\rm{ext}} \nonumber \\
    & - \langle \bar \nabla_{I} \cM_{\bar p r}\bar\psi^{\bar p} \psi^{r}\rangle_{\rm{ext}}  
\end{align}
where the notation is defined in fig.~\ref{fig:O_ext}.
\begin{figure}
    \centering

    \begin{tikzpicture}
            \coordinate (d) at (-3.75, 0);
            \coordinate (c) at (-3, 0);	
	    	\coordinate (a) at (-2, 1);
		        \node[left]  at (d) {$\langle {\cal O} \rangle_{\rm ext} \quad = \quad \mathlarger{ \sum_{a=1}^n}$};
	        \node[left]  at (c) {$a$};
	    	
	    	\coordinate (f) at (1, 1);
	    	\coordinate (l) at (1, -1);
	    	
	    	\coordinate (m4) at (-2,0);
	    	
	    	\coordinate (mn) at (-0.25, 0);
	    
	    	\coordinate (d1) at (1.00, 0);
	    	\coordinate (d2) at (0.85, 0.5);
	    	\coordinate (d3) at (0.85, -0.5);
	    	
	    	\draw [hard] (c) -- (m4);
	    	\draw [hard] (mn) -- (m4);
	    	
	    	\draw [hard] (f) -- (mn);
	        \draw [hard] (l) -- (mn);
	    	
            \draw[fill=lightgray, opacity=1] (mn) circle (0.65);
            	\node at (mn) {$A_{n}$};
            
            \draw[fill=white, opacity=1] (m4) circle (0.2);
            \node[below]  at (-2,-0.2) {${\cal O}(q) $};
	    	
            \draw[fill=black, opacity=1] (d1) circle (0.02);
            \draw[fill=black, opacity=1] (d2) circle (0.02);
            \draw[fill=black, opacity=1] (d3) circle (0.02);
\end{tikzpicture}
    \caption{Diagrams computing $\langle {\cal O}\rangle_{\rm ext}$, which sums over the insertion of an operator ${\cal O}$ on each external leg $a$ of the $n$-particle amplitude. This figure is directly reproduced from ref.~\cite{Cheung:2021yog}.}
    \label{fig:O_ext}
\end{figure}
By evaluating these operator insertions, we find that the first line in \cref{eq:fermion_operator_insertion} either vanishes due to the on-shell condition, or it cancels a propagator and becomes a local term multiplying the amplitude. These local terms are $-\bar \Gamma^{p}_{I  r}$ or $-\bar \Gamma^{\bar p}_{I \bar r}$, depending on whether the operator is inserted on an incoming fermion or anti-fermion line. This is precisely the fermion connections in \cref{eq:fermion_Christoffel1,eq:fermion_Christoffel2}, and when combined with the scalar contributions, they complete the covariant derivative $\bar \nabla_{I}$ in the soft theorem in \cref{eq:soft_theorem_fermion}. 
What is left is the insertion of the operator in the second line in \cref{eq:fermion_operator_insertion}, which becomes the second and third lines of the soft theorem in \cref{eq:soft_theorem_fermion}.

In \cref{sec:examples} we will check some examples of the soft theorem for an effective field theory with scalars and fermions.

\subsection{Gauge bosons}
\label{sec:gauge}

The last particle to make an appearance is the gauge boson. In this case, we use the geometric construction in \cref{eq:gauge_metric}. Due to the block-diagonal structure of the metric, the fermions and gauge bosons do not couple directly through the geometry, and we can simply ignore the fermions for the moment. 

The soft theorem for a massless scalar in a theory with scalars and gauge bosons is
\begin{align}
    \label{eq:soft_theorem_gauge}
    \lim_{q\rightarrow 0} A_{n+1,\alpha_1 \cdots \alpha_{n} j} =& \bar \nabla_{j} A_{n,\alpha_1 \cdots \alpha_n} + \sum_{a\in \{\textrm{scalar}\}} \frac{\nabla_{j} V_{i_a}^{\;\; j_a}}{(p_a+q)^2 - m^2_{j_a}} \left(1 + q^{\mu} \frac{\partial}{\partial p_a^{\mu}} \right) A_{n,\alpha_1 \cdots j_a \cdots \alpha_n} \nonumber \\
    & + \sum_{a\in \{\textrm{scalar}\}} \sum_{\rm{spin}} i\frac{(2\nabla_j t_{i_a \, B}) (\epsilon^{Bb*}\cdot q)}{(p_{a}+q)^2 - m^2_b}A_{n,\alpha_1\cdots b \cdots \alpha_{n}} \nonumber \\
    & + \sum_{b\in \{\textrm{gauge}\}} i\frac{(2\nabla_j t^{j_a}_{\,B})  ( \epsilon^{B}_{b} \cdot q)}{(p_b+q)^2 - m^2_{j_a}}A_{n,\alpha_1\cdots j_a \cdots \alpha_{n}} \nonumber \\
    & - \sum_{b\in \{\textrm{gauge}\}} \sum_{\rm{spin}} \frac{(\bar \nabla_{j} m_{BA}^{2})  ( \epsilon^{B}_{b} \cdot \epsilon^{Aa*})}{(p_b+q)^2 - m^2_{a}}\left(1 + q^{\mu} \frac{\partial}{\partial p_a^{\mu}} \right) A_{n,\alpha_1\cdots a \cdots \alpha_{n}} \,.
\end{align}
Note that the covariant derivative $\bar \nabla_{i}$ now uses the connections derived from the metric in \cref{eq:gauge_metric} and sees the gauge group indices of the gauge bosons. Here, the spin is summed over for the external gauge-boson polarization vector in the $n$-point amplitude and the polarization vector in the prefactors $\epsilon^{Aa*}(p+q)$, evaluated at shifted momentum.

As a first check, we act the soft theorem on the on-shell condition for the gauge boson,
\begin{align}
    (p^2_{b} g_{AB} - m_{A B}^{2} ) \epsilon^{Bb}_{\mu} = 0 . 
\end{align}
The covariant derivative picks up the variation of the mass matrix,
\begin{align}
    \label{eq:gauge_onshell1}
   \bar \nabla_{j} (p^2_{b} g_{AB} - m_{A B}^{2} ) \epsilon^{Bb}_{\mu} = - (\bar \nabla_{j} m_{A B}^{2} ) \epsilon^{Bb}_{\mu}  . 
\end{align}
The first term vanishes due to metric compatibility, $\bar \nabla g = 0$. 

Then we contract the on-shell condition with the last term of the soft theorem,
\begin{align}
    \label{eq:gauge_onshell2}
   &-\sum_{\rm{spin}} \frac{(\bar \nabla_{j} m_{BC}^{2})  ( \epsilon^{B}_{b} \cdot \epsilon^{Cc*})}{(p_b+q)^2 - m^2_{c}}\left(1 + q^{\mu} \frac{\partial}{\partial p_b^{\mu}} \right)  (p^2_{b} g_{AC^\prime} - m_{AC^\prime}^{2} ) \epsilon^{C^\prime}_{c\mu}
   \nonumber \\ 
   &=
    -\sum_{\rm{spin}} \frac{(\bar \nabla_{j} m_{BC}^{2})  ( \epsilon^{B}_{b} \cdot \epsilon^{Cc*})}{(p_b+q)^2 - m^2_{c}}  ((p_{b}+q)^2 g_{AC^\prime} - m_{AC^\prime}^{2} ) \epsilon^{C^\prime}_{c\mu}
    =
    +(\bar \nabla_{j} m_{BA}^{2})   \epsilon^{B}_{b\mu} .
\end{align}
We see that \cref{eq:gauge_onshell1,eq:gauge_onshell2} cancel, which means that the soft theorem commutes with the on-shell conditions. 

Note that there are no terms of the form $\nabla m^2$ in the soft theorem for massless gauge bosons since 
\begin{align}
    \nabla_K m_{ab}^{2}(v) = \nabla_K \left(t^I{}_{a}(v)\, t_{Ib}(v)\right) = \nabla_K t^I{}_{a}(v)\, t_{Ib}(v) + t^I{}_a(v) \nabla_K  t_{Ib}(v),
    \label{eq:dmzero}
\end{align}
and in the unbroken phase, one has $t^{I}_{a}(v)=0$, so
\begin{equation}
    \nabla_K m_{ab}^{2}(v) = 0 .
\end{equation}
This means that the gauge boson masses vanish even in an infinitesimal neighborhood of the unbroken VEV.

To fully understand the form of the soft theorem, we need to consider the interplay between longitudinal gauge bosons and Goldstone bosons. We will show how different representations of the soft theorem are linked via the Goldstone boson equivalence theorem. Consider the second and third lines in \cref{eq:soft_theorem_gauge}. In unitary gauge, the Goldstone boson decouples, and we only exchange massive gauge bosons
\begin{align}
    \label{eq:NGBequivalence1}
    & + i\frac{(2\nabla_j t_{i_a}^{b}) ( q^{\mu})}{2 p_a \cdot q} (-1)\left( \eta^{\mu\nu} - \frac{p^\mu_a p^\nu_a}{m_a^2} \right)  A_{n,\alpha_1\cdots b\nu \cdots \alpha_{n}} .
\end{align}

In $R_\xi$ gauge, we instead get 
\begin{align}
    \label{eq:NGBequivalence2}
    & + i\frac{(2\nabla_j t_{i_a}^{b}) (q^\mu)}{(p_{a}+q)^2 - m^2_a} (-1) \left(\eta^{\mu\nu} - (1-\xi) \frac{p^\mu_a p^\nu_a}{p^2_a - \xi m^2_a} \right) A_{n,\alpha_1\cdots b\nu \cdots \alpha_{n}} \nonumber \\
    & + i\frac{(2\nabla_j t^{j_a}_{\,B})  ( \epsilon^{B}_{b} \cdot q)}{(p_b+q)^2 - \xi m^2_{a}}A_{n,\alpha_1\cdots j_a \cdots \alpha_{n}} \nonumber \\
    =& + i\frac{(2\nabla_j t_{i_a}^{b}) (q^\mu)}{2 p_{a}\cdot q} (-1) \left(\eta^{\mu\nu} -  \frac{p^\mu_a p^\nu_a}{ m^2_a} \right) A_{n,\alpha_1\cdots b\nu \cdots \alpha_{n}} \nonumber \\
    & + i\frac{(2\nabla_j t_{i_a}^{b}) (q^\mu)}{(p_{a}+q)^2 - \xi m^2_a} (-1) \left( \frac{p^\mu_a p^\nu_a}{ m^2_a} \right) A_{n,\alpha_1\cdots b\nu \cdots \alpha_{n}} \nonumber \\
    & + i\frac{(2\nabla_j t^{j_a}_{\,B})  ( \epsilon^{B}_{b} \cdot q)}{(p_b+q)^2 - \xi m^2_{a}}A_{n,\alpha_1\cdots j_a \cdots \alpha_{n}} .
\end{align}
If we identify the longitudinal polarization with the momentum $\epsilon^\mu_{L} \rightarrow p^\mu/m$, then the last two terms in \cref{eq:NGBequivalence2} cancel, and we end up with the same result as in \cref{eq:NGBequivalence1}. This is due to the Goldstone boson equivalence theorem in \cref{eq:NBGequiv}. Here, instead of taking the high-energy limit for the longitudinal gauge boson, we take the soft limit for a scalar. These limits yield the same result because the longitudinal gauge boson has a large energy relative to the soft scalar.

Incidentally, if we instead used the geometry-kinematics map, the soft theorem would take the form
\begin{align}
    \lim_{q\rightarrow 0} A_{n+1,\alpha_1 \cdots \alpha_{n} j} =& \nabla^{\prime}_{j} A_{n,\alpha_1 \cdots \alpha_n} . 
\end{align}
This coincides with the first term in \cref{eq:soft_theorem_gauge}, but the greater freedom in the mapping also puts the nonlocal terms into the covariant derivative as extensions of the connection. 
Ref.~\cite{Cheung:2022vnd} showed that this soft theorem also captures the leading and subleading soft photon theorem. 

The proof of the geometric soft theorem with gauge bosons is completely analogous to that for scalars and vectors, so we will not describe it here. Instead, we will directly check the soft theorem in various examples in section~\ref{sec:examples}.



\section{\label{sec:double-soft-theorem}Double Soft Theorems}

Another way to study scattering amplitudes is to send the momenta of multiple particles to zero. If we do so in a consecutive order, we simply need to apply the geometric soft theorem multiple times. However, if the momenta are sent to zero simultaneously, we will discover a genuinely new geometric structure in the scattering amplitudes: the curvature. This demonstrates the non-abelian nature of pion scattering \cite{Weinberg:1966kf}. 

To ease the presentation, we turn off all couplings which appear in the nonlocal terms in the geometric soft theorem, i.e., the scalar potential $V(\phi)$, the fermion mass matrix ${\cal M}(\phi)$, and we make the particles neutral, i.e., $t = 0$ and $\nabla t = 0$. This avoids multiple soft poles in the expressions. 

First, we consider the double soft limit where the momenta of two scalars are taken to zero at the same rate. This will be an extension of the double soft theorem in a scalar effective field theory \cite{Cheung:2021yog}. Then, we change the protagonists and consider the double soft limit where the momenta of two fermions of opposite helicity are sent to zero at the same rate. This new double soft fermion theorem has striking similarities to the double soft scalar theorem. Here, all momenta are outgoing.

\subsection{Scalars}

The double scalar soft theorem is identical to the form derived in ref.~\cite{Cheung:2021yog}, when using the appropriate geometric extensions when fermions and gauge bosons are present. Here, the potential and other terms singular in the soft limit are neglected. The simultaneous double soft theorem is
\begin{align}
    \label{eq:double_scalar_soft_theorem}
    \lim_{q_1,q_2\rightarrow 0} A_{n+2,\alpha_1\cdots \alpha_{n} j_1 j_2} =& \bar \nabla_{(j_1}\bar \nabla_{j_2)} A_{n,\alpha_1 \cdots \alpha_n} + \frac{1}{2} \sum^{n}_{a=1} \frac{p_a \cdot (q_1 - q_2) }{p_a \cdot (q_1 + q_2)}\bar R_{j_1 j_2 \;\;\;\;\alpha_a}^{\quad\; \beta_a} \;A_{n,\alpha_1 \cdots \beta_a \cdots \alpha_n} .
\end{align}
The particles with flavor labels $\{\alpha_1,\cdots,\alpha_n\}$ can be any combination of massless scalars, fermions, or gauge bosons. Remarkably, the same double soft theorem holds regardless of whether the scalars couple to fermions, gauge bosons, or other scalars; the various interactions are captured by the combined curvature $\bar R_{j_1 j_2 \alpha_a \beta_a}$.
We present several examples of soft limits of scattering amplitudes in \cref{sec:examples}, where the double scalar soft theorem can be checked.

\subsection{Fermions}

We can also consider the soft limit of two fermions with opposite helicities. For convenience, let us use the spinor-helicity formalism (following the conventions in ref.~\cite{Dixon:1996wi}). The result is
\begin{align}
    \label{eq:double_fermion_soft_theorem}
    \lim_{q_1,q_2\rightarrow 0} A_{n+2,\alpha_1\cdots \alpha_{n} \bar r_1 r_2} =& \frac{1}{2} \left\{ \lim_{q_1\rightarrow 0}, \lim_{q_2\rightarrow 0}\right\} A_{n+2,\alpha_1 \cdots \alpha_n \bar r_1 r_2} 
    \nonumber \\ &
    + \frac{1}{2} \sum^{n}_{a=1} \frac{[ q_1 |p_a | q_2 \rangle }{p_a \cdot (q_1 + q_2)} \bar R_{\bar r_1 r_2 \;\;\;\;\alpha_a}^{\quad\; \beta_a} \;A_{n,\alpha_1 \cdots \beta_a \cdots \alpha_n} .
\end{align}
The double fermion soft theorem is equal to the double scalar soft theorem under the replacement $p_a \cdot (q_1 - q_2) \rightarrow [q_1| p_a |q_2\rangle$, as first noted in ref.~\cite{Chen:2014xoa} for supersymmetric theories. The proof is diagrammatic and the same as for  two soft scalars (see ref.~\cite[sec.~6.2]{Cheung:2021yog}). The first term in \cref{eq:double_fermion_soft_theorem} is written in terms of the anticommutator of two consecutive soft limits, rather than as covariant derivatives acting on the lower-point amplitude. This is because we do not have a geometric way of writing the soft limit of a single fermion in terms of lower-point amplitudes. However, the single soft fermion limit vanishes in many instances, and then we end up with the simpler form of \cref{eq:double_fermion_soft_theorem},
\begin{align}
    \label{eq:double_fermion_soft_theorem_simple}
    \lim_{q_1,q_2\rightarrow 0} A_{n+2,\alpha_1\cdots \alpha_{n} \bar r_1 r_2} =&  \frac{1}{2} \sum^{n}_{a=1} \frac{[ q_1 |p_a | q_2 \rangle }{p_a \cdot (q_1 + q_2)} \bar R_{\bar r_1 r_2 \;\;\;\;\alpha_a}^{\quad\; \beta_a} \;A_{n,\alpha_1 \cdots \beta_a \cdots \alpha_n} .
\end{align}

The curvature that enters the double soft theorem depends on the other particles in the theory. In the presence of scalar particles, the mixed scalar-fermion curvature $\bar R_{\bar r_1 r_2 i j}$ controls the nonlocal term, whose expression is given in \cref{eq:scalar-fermion_curvature}. For fermion-fermion interactions, the four-fermion curvature $\bar R_{\bar r_1 r_2 \bar r_3 r_4}$ in \cref{eq:fermion-fermion_curvature} enters the double soft theorem. Even though the kinematic expressions that come with these different curvature components in scattering amplitudes are very different, they reduce to the exact same term in the double soft limit.

\subsection{Other simultaneous soft limits}

Now, the door is open to consider even more exotic simultaneous soft limits. Take, as an example, the simultaneous soft limit of one scalar and one fermion. Concretely, we take a positive-helicity fermion with a holomorphic soft scaling, in spinor-helicity variables $(\lambda, \tilde \lambda) \to (z \lambda, \tilde \lambda)$, with $z$ small. Rather than deriving the double soft theorem as we did above, let us try to guess the answer from the intuition we have accrued. A natural guess for the double soft limit is
\begin{align}
    \label{eq:double_scalar_fermion_soft_theorem}
    \lim_{q_1,q_2\rightarrow 0} A_{n+2,\alpha_1\cdots \alpha_{n} \bar r_1 j_2} =&  \lim_{q_1\rightarrow 0}\bar \nabla_{j_2} A_{n+1,\alpha_1 \cdots \alpha_n \bar r_1} + \frac{1}{2} \sum^{n}_{a \in \{\textrm{fermion}\}} \frac{[ q_1 | q_2 | p_a \rangle }{p_a \cdot (q_1 + q_2)} \bar R_{\bar r_1 r_a \;\;\; j_2}^{\quad\;\; j_a} \;A_{n,\alpha_1 \cdots j_a \cdots \alpha_n} \nonumber \\ &
    + \frac{1}{2} \sum_{a\in \{\textrm{scalar}\}} \frac{[ q_1 | q_2 | p_a \rangle }{p_a \cdot (q_1 + q_2)} \bar R_{\bar r_1  \;\;\; i_a j_2}^{\;\;\; \bar s_a} \;A_{n,\alpha_1 \cdots \bar s_a \cdots \alpha_n} \,.
\end{align}
This mixed double soft theorem can be derived via a diagrammatic approach analogous to the double scalar and fermion soft theorems. 
We have verified in the examples in \cref{sec:examples} that this mixed double soft theorem indeed holds. Again, the soft theorem is identical in form to the double scalar and double fermion soft theorem, up to a simple replacement of a kinematic factor. 
The first term in \cref{eq:double_scalar_fermion_soft_theorem} is written in terms of the single soft limit of a lower-point amplitude rather than as $\bar\nabla_{(\bar r_1}\bar\nabla_{i_2)} A_{n,\alpha_1 \cdots \alpha_n}$. The single soft limit of a fermion is hard to interpret in terms of scattering amplitudes because it would involve the derivative of a would-be amplitude with an odd number of fermions\footnote{The exception is for supersymmetric theories, where the soft fermion theorem is related to a soft boson theorem (e.g., for Goldstinos \cite{Uematsu:1981rj} or photinos \cite{Dumitrescu:2015fej}).}.

Based on this, we expect that \textit{any} double soft limit will be universal. It will involve the curvature in field space, accompanied by an appropriate kinematic factor to account for the helicity weight of the soft particles. Indeed, we know that this is true for double soft limits involving gauge bosons, through the geometry-kinematics map. In this case, the double soft limit will be identical to \cref{eq:double_scalar_soft_theorem}, with replacements $\nabla \rightarrow \nabla^{\prime}$ and $R \rightarrow R^{\prime}$. The kinematic factors that carry the helicity weights are folded into the geometry, which now also depends on the kinematics.

One can also consider more soft particles. The case with three soft scalars was analyzed in ref.~\cite{Cheung:2021yog}. As one might have guessed, the triple soft theorem involves various terms with $\nabla^3$, $R \nabla$, and $\nabla R$ acting on the lower-point amplitude. We expect that the generalization of multiple soft limits with a mixture of particles will be the natural generalizations of the scalar case, but where the kinematic factors are replaced and the geometry is extended. We will not explore this direction further here.


\section{Examples}
\label{sec:examples}

We now present tree-level scattering amplitudes for scalars, fermions, and gauge bosons. With these amplitudes, we can check the single and double soft theorems. All momenta are outgoing, and we use the spinor-helicity conventions for massless and massive particles in refs.~\cite{Dixon:1996wi,Arkani-Hamed:2017jhn,Durieux:2019eor}.

\subsection{Scalars}
\label{sec:example_scalar}

We start by listing some scattering amplitudes for scalars with two-derivative interactions. The corresponding Lagrangian is
\begin{align}
    \Lagr = \frac{1}{2} h_{IJ}(\phi) (\partial_{\mu} \phi)^{I} (\partial^{\mu} \phi)^{J} . 
\end{align}
The scattering amplitudes for four and five particles are
\begin{align}
    \label{eq:amplitude4_scalar}
    A_{4,i_1 i_2 i_3 i_4} =& R_{i_1 i_3 i_2 i_4} s_{34} + R_{i_1 i_2 i_3 i_4} s_{24} , \\
    \label{eq:amplitude5_scalar}
    A_{5,i_1 i_2 i_3 i_4 i_5} =& \nabla_{i_3} R_{i_1 i_4 i_2 i_5} s_{45} + \nabla_{i_4} R_{i_1 i_3 i_2 i_5} s_{35} + \nabla_{i_4} R_{i_1 i_2 i_3 i_5} s_{25} \nonumber \\ 
    &+ \nabla_{i_5} R_{i_1 i_3 i_2 i_4} s_{34} + \nabla_{i_5} R_{i_1 i_2 i_3 i_4} (s_{24} + s_{45}) , 
\end{align}
where $s_{ij} = (p_{i} + p_{j})^2$. 
We will use these amplitudes to illustrate the geometric soft theorem for scalar effective field theories in \cref{eq:soft_theorem_scalar}. For more examples and the original derivation, see ref.~\cite{Cheung:2021yog}.

Take the limit $p_4\rightarrow 0$ of \cref{eq:amplitude4_scalar},
\begin{align}
    \lim_{p_4 \rightarrow 0} A_{4,i_1 i_2 i_3 i_4} = 0.
\end{align}
This adheres to the geometric soft theorem, because the scalar three-particle amplitude is zero when the potential is absent. 

Next, look at the limit $p_5 \rightarrow 0$ of \cref{eq:amplitude5_scalar}:
\begin{align}
    \lim_{p_5 \rightarrow 0} A_{5,i_1 i_2 i_3 i_4 i_5} = \nabla_{i_5} R_{i_1 i_3 i_2 i_4} s_{34} + \nabla_{i_5} R_{i_1 i_2 i_3 i_4} s_{24} = \nabla_{i_5} A_{4,i_1 i_2 i_3 i_4} .
\end{align}
This is precisely the statement of the geometric soft theorem with no potential: the soft limit of the amplitude is equal to the covariant derivative acting on the lower-point amplitude. This is the cleanest illustration of the geometric soft theorem. However, the soft theorem is valid for general scalar effective field theories, including potential and higher-derivative interactions. Additional examples can be found in Ref.~\cite{Cheung:2021yog}.

\subsection{Fermions}

Next, we look at scattering amplitudes with fermions and scalars, coming from the one-derivative fermion bilinear operators and scalar operators with two derivatives as well as from the four-fermion operators. These are the operators which appear in the scalar-fermion metric in \cref{eq:fermion_metric}. The Lagrangian is
\begin{align}
    \label{eq:Lagrangian_fermion_example1}
    \Lagr =& \frac{1}{2} h_{IJ}(\phi) (\partial_{\mu} \phi^{I}) (\partial^{\mu} \phi^{J}) 
    + i \frac{1}{2} k_{\bar p r}(\phi) (\bar \psi^{\bar p} \overset{\leftrightarrow}{\slashed \partial} \psi^{r}) + i \omega_{\bar p r I}(\phi) (\bar \psi^{\bar p} \gamma_{\mu} \psi^{r}) (\partial^{\mu} \phi^I) \\ 
    &+ c_{\bar p r \bar s t}(\phi) (\bar \psi^{\bar p} \gamma_{\mu} \psi^{r}) (\bar \psi^{\bar s} \gamma^{\mu} \psi^{t}). \nonumber 
\end{align}
The scattering amplitude with two fermions and one scalar vanishes. The scattering amplitudes with two, three, or four scalars are
\begin{align}
    \label{eq:amplitude4_fermion}
    A_{4,\bar r_1 r_2 i_3 i_4} =& -[1|p_4|2\rangle \bar R_{\bar r_1 r_2 i_3 i_4} , \\
    \label{eq:amplitude5_fermion}
    A_{5,\bar r_1 r_2 i_3 i_4 i_5} =& -[1|p_4|2\rangle \bar \nabla_{i_5} \bar R_{\bar r_1 r_2 i_3 i_4} - [1|p_5|2\rangle \bar \nabla_{i_4} \bar R_{\bar r_1 r_2 i_3 i_5} , \\
    \label{eq:amplitude6_fermion}
    A_{6,\bar r_1 r_2 i_3 i_4 i_5 i_6} =& \left\{ -[1|p_4|2\rangle \left( \frac{1}{4} \bar \nabla_{i_5} \bar \nabla_{i_6} \bar R_{\bar r_1 r_2 i_3 i_4} + \frac{1}{6} \bar R_{\bar r_1 r_2 i_3}^{\quad\quad\; j} \bar R_{i_4 i_5 i_6 j} - \frac{1}{4} \bar R_{\bar r_1 s i_3 i_5} \bar R^{s}_{\;\; r_2 i_4 i_6} \right.\right. \nonumber \\ &  \qquad\qquad\qquad - (3\leftrightarrow 4) + (5\leftrightarrow 6) - (3\leftrightarrow 4, 5\leftrightarrow 6) \bigg) + {\rm cycl(4,5,6)} \bigg\} \nonumber \\
    &+ \left\{ \frac{[1|(p_5-p_6)P_{234}(p_3-p_4)| 2\rangle}{16 s_{234}} \bar R_{\bar r_1 s i_5 i_6} \bar R^{s}_{\;\; r_2 i_3 i_4} + \textrm{perm}(3,4,5,6) \right\} \nonumber \\
    & + \left\{ \frac{[1|p_6|2\rangle}{3 s_{345}} \bar R_{\bar r_1 r_2 i_6}^{\quad\quad\; j} \big[ s_{34} (\bar R_{i_3 i_4 i_5 j} - 2 \bar R_{i_3 i_5 i_4 j} ) + s_{35} (\bar R_{i_3 i_5 i_4 j} - 2 \bar R_{i_3 i_4 i_5 j} ) \right. \nonumber \\ &\left. \qquad\qquad\qquad\qquad\qquad + s_{45} (\bar R_{i_3 i_4 i_5 j} + \bar R_{i_3 i_5 i_4 j} ) \right] + \textrm{cycl}(3,4,5,6) \bigg\}  .
\end{align}
Here, $s_{ijk} = (p_{i} + p_{j} + p_{k})^2$ and $P_{ijk}^{\mu} = p^{\mu}_{i} + p^{\mu}_{j} + p^{\mu}_{k}$, and we sum over all or cyclic permutations denoted by perm() or cycl().

The scattering amplitudes with four or six fermions but no scalars are 
\begin{align}
    \label{eq:amplitude4_fermion-fermion}
    A_{4,\bar r_1 r_2 \bar r_3 r_4} =& [13] \langle 42 \rangle  \bar R_{\bar r_1 r_2 \bar r_3 r_4} ,  \\
    \label{eq:amplitude6_fermion-fermion}
    A_{6,\bar r_1 r_2 \bar r_3 r_4 \bar r_5 r_6} =& \left(-\frac{[5|(p_1+p_3)|2\rangle [13]\langle 64 \rangle}{s_{123}}\bar R_{\bar r_1 r_2 \bar r_3}{}{}^{\bar s}\bar R_{\bar s r_4 \bar r_5 r_6}+\textrm{cycl}(1,3,5)\right) +\textrm{cycl}(2,4,6) . 
\end{align}

Let us now check the new geometric soft theorem in the presence of fermions. First, the limit $p_4 \rightarrow 0$ for the two scalar, two fermion amplitude is
\begin{align}
    \lim_{p_4 \rightarrow 0} A_{4,\bar r_1 r_2 i_3 i_4} = 0,
\end{align}
which is consistent with the soft theorem in \cref{eq:soft_theorem_fermion}.

A more nontrivial example is the $p_5 \rightarrow 0$ soft limit of the five-particle amplitude,
\begin{align}
    \lim_{p_5 \rightarrow 0} A_{5,\bar r_1 r_2 i_3 i_4 i_5} = -[1|p_4|2\rangle \bar \nabla_{i_5} \bar R_{\bar r_1 r_2 i_3 i_4} = \bar \nabla_{i_5} A_{4,\bar r_1 r_2 i_3 i_4} .
\end{align}
This is the scalar-fermion soft theorem in \cref{eq:soft_theorem_fermion} with the potential and fermion mass matrix turned off. Structurally, it is identical to the geometric soft theorem for scalars, but with the crucial difference that $\nabla_i \rightarrow \bar \nabla_i$. The geometric soft theorem depends on the combined scalar-fermion geometry dictated by the metric in \cref{eq:fermion_metric}. 

Next, consider the limit $p_6 \rightarrow 0$ of the six-particle amplitude in \cref{eq:amplitude6_fermion},
\begin{align}
    \lim_{p_6 \rightarrow 0} A_{6,\bar r_1 r_2 i_3 i_4 i_5 i_6} = -[1|p_4|2\rangle \bar \nabla_{i_6} \bar \nabla_{i_5} \bar R_{\bar r_1 r_2 i_3 i_4} - [1|p_5|2\rangle \bar \nabla_{i_6} \bar \nabla_{i_4} \bar R_{\bar r_1 r_2 i_3 i_5} = \bar \nabla_{i_6} A_{5,\bar r_1 r_2 i_3 i_4 i_5} . 
\end{align}
This example showcases an intricate cancellation between local $R^2$ terms and $R^2$ terms with factorization channels which become localized in the soft limit.

We can also study these amplitudes in the double soft limit. Take the simultaneous soft limit $p_5,p_6 \rightarrow 0$ of the six-particle amplitude in \cref{eq:amplitude6_fermion},
\begin{align}
    \lim_{p_5,p_6 \rightarrow 0} A_{6,\bar r_1 r_2 i_3 i_4 i_5 i_6} =& -[1|p_4|2\rangle \bar \nabla_{(i_5} \bar \nabla_{i_6)} \bar R_{\bar r_1 r_2 i_3 i_4} \nonumber \\
    &+ \frac{(s_{15} - s_{16}) \bar R_{i_5 i_6 \bar r_1}^{\quad\quad \bar s}}{2 (s_{15} + s_{16})} \; \left( -[1|p_4| 2\rangle \bar R_{s r_2 i_3 i_4}  \right) \nonumber \\
    &+ \frac{(s_{25} - s_{26}) \bar R_{ i_5 i_6 r_2}^{\quad\quad s}}{2 (s_{25} + s_{26})} \; \left( -[1|p_4| 2\rangle\bar R_{\bar r_1 s i_3 i_4} \right)  \nonumber \\
    & - \frac{( s_{35} - s_{36} ) \bar R_{i_5 i_6 i_3}^{\quad\quad j}}{2 (s_{35} + s_{36})}  \; \left( -[1|p_4|2\rangle  \bar R_{\bar r_1 r_2 j i_4} \right) \nonumber \\
    & - \frac{(s_{45} - s_{46}) \bar R_{i_5 i_6 i_4}^{\quad\quad j}}{2 (s_{45} + s_{46})} \;  \left( -[1|p_4|2\rangle \bar R_{\bar r_1 r_2 i_3 j}  \right) \nonumber \\
    &= \bar \nabla_{(i_5} \bar \nabla_{i_6)} A_{4,\bar r_{1} r_2 i_3 i_4} + \frac{1}{2}\sum_{a=1}^{4} \frac{p_{a}\cdot (p_{5} - p_{6})}{p_{a}\cdot (p_{5} + p_{6})} \bar R_{i_5 i_6 \;\;\;\;\alpha_{a}}^{\quad\;\beta_{a}} \;A_{4,\dots \beta_{a} \dots} \,.
\end{align}
This novel double soft theorem is again structurally similar to the corresponding double soft theorem for scalar theories, but with the uplifts $\nabla_{i} \rightarrow \bar \nabla_{i}$ and $R \rightarrow \bar R$.

With these amplitudes in hand, we can ask a different question. What happens when the momenta of two fermions are sent to zero? Take two fermions with opposite helicity and democratically scale their spinors in the soft limit. The double fermion soft limit of the six-particle amplitude in \cref{eq:amplitude6_fermion} is
\begin{align}
    \lim_{p_1,p_2\rightarrow 0} A_{6,\bar r_1 r_2 i_3 i_4 i_5 i_6} =& \frac{1}{2}\frac{[1|p_6|2\rangle}{ p_{6}\cdot(p_{1} + p_{2})} \bar R_{\bar r_1 r_2 \;\; i_6}^{\quad\;\; j}\; A_{4,i_3 i_4 i_5 j} + \textrm{cycl}(3,4,5,6) .
\end{align}
This agrees with \cref{eq:double_fermion_soft_theorem_simple}, since the single soft fermion limit vanishes. 

As a last example, we take the double fermion soft limit of the six-fermion amplitude in \cref{eq:amplitude6_fermion-fermion}, which gives
\begin{align}
    \lim_{p_1,p_2 \rightarrow 0} A_{6,\bar r_1 r_2 \bar r_3 r_4 \bar r_5 r_6} =
    & \frac{[1 |p_3|2\rangle \bar R_{\bar r_1 r_2 \bar r_3}^{\quad\quad \bar s}}{s_{13} + s_{23}} \;  \left( [35]\langle 64\rangle \bar R_{\bar s r_4 \bar r_5 r_6}  \right) \nonumber \\
    & + \frac{[1 |p_4|2\rangle \bar R_{\bar r_1 r_2 r_4}^{\quad\quad s}}{ s_{14} + s_{24}} \;  \left( [35]\langle 64\rangle \bar R_{\bar r_3 s \bar r_5 r_6}  \right) \nonumber \\ 
    & + \frac{[1 |p_5|2\rangle \bar R_{\bar r_1 r_2 \bar r_5}^{\quad\quad \bar s}}{ s_{15} + s_{25}} \;  \left( [35]\langle 64\rangle \bar R_{\bar r_3 r_4 \bar s r_6}  \right) \nonumber \\
    & + \frac{[1 |p_6|2\rangle \bar R_{\bar r_1 r_2 r_6}^{\quad\quad s}}{s_{16} + s_{26}} \;  \left( [35]\langle 64\rangle \bar R_{\bar r_3 r_4 \bar r_5 s}  \right) .
\end{align}
Again, this agrees with \cref{eq:double_fermion_soft_theorem_simple}.

\subsection{Gauge bosons}

Third, we consider the scattering of scalars and gauge bosons. For the sake of illustration, we take the scalars to be neutral and massless. The relevant Lagrangian is
\begin{align}
    \Lagr =& \frac{1}{2} h_{IJ}(\phi) (\partial_{\mu} \phi)^{I} (\partial^{\mu} \phi)^{J} - \frac{1}{4} g_{AB}(\phi) F_{\mu\nu}^{A} F^{B\mu\nu} .
\end{align}
The scattering amplitudes for two positive-helicity gauge bosons and one, two, or three scalars are
\begin{align}
    A_{3,a_1 a_2 i_3} =& [12]^2 \frac{1}{2} \nabla_{i_3} g_{a_1 a_2} , \\
    A_{4,a_1 a_2 i_3 i_4} =& [12]^2 \frac{1}{2} \bar \nabla_{i_4} \nabla_{i_3} g_{a_1 a_2} , \\
    A_{5,a_1 a_2 i_3 i_4 i_5} =& [12]^2 \frac{1}{2} \bar \nabla_{(i_4} \bar \nabla_{i_5} \nabla_{i_3)} g_{a_1 a_2} \nonumber \\
    & +\left\{ \frac{(\nabla_{i_5} g_{a_1 b_1}) g^{b_1 b_2} (\nabla_{i_4} g_{b_2 b_3})g^{b_3 b_4} (\nabla_{i_3} g_{b_4 a_2})}{s_{15}s_{23}}\left[ \frac{1}{8} [1|p_5 p_3|2]^2 - \frac{1}{24} s_{15} s_{23} [12]^2\right]  
    \right.\nonumber \\ & \qquad\qquad
    + \textrm{perm}(3,4,5)\Big\}\nonumber\\
    & + \frac{[12]^2 \nabla^{j} g_{a_1 a_2}}{6 s_{345}} \left[ s_{34} (\bar R_{i_3 i_4 i_5 j} - 2 \bar R_{i_3 i_5 i_4 j} )  + s_{35} (\bar R_{i_3 i_5 i_4 j} - 2 \bar R_{i_3 i_4 i_5 j} ) 
    \right. \nonumber\\&\left.\qquad\qquad\qquad\quad
    + s_{45} (\bar R_{i_3 i_4 i_5 j} + \bar R_{i_3 i_5 i_4 j} ) \right] . 
\end{align}
Note that the amplitudes do not vanish due to metric compatibility, $\bar \nabla g = 0$, because the connection in the covariant derivative $\nabla_i$ is for the scalar bundle, i.e., $\nabla_{i} g_{ab} = g_{ab,i}$. However, for the four-particle amplitude the connection for the full scalar--gauge-boson geometry is in play:
\begin{align}
    \bar \nabla_{i_4} \nabla_{i_3} g_{a_1 a_2} = \nabla_{i_4} \nabla_{i_3} g_{a_1 a_2} - \bar \Gamma^{b}_{a_1 i_4} \nabla_{i_3} g_{b a_2} - \bar \Gamma^{b}_{a_2 i_4} \nabla_{i_3} g_{a_1 b} .
\end{align}

Now we can study the single soft limit, starting with $p_4 \rightarrow 0$ in the four-particle amplitude,
\begin{align}
    \lim_{p_4 \rightarrow 0} A_{4,a_1 a_2 i_3 i_4} =& [12]^2 \frac{1}{2} \bar \nabla_{i_4} \nabla_{i_3} g_{a_1 a_2} = \bar \nabla_{i_4} A_{3,a_1 a_2 i_3} .
\end{align}
A more involved example is the soft limit $p_5 \rightarrow 0$ for the five-particle amplitude,
\begin{align}
    \lim_{p_5 \rightarrow 0} A_{5,a_1 a_2 i_3 i_4 i_5} =& [12]^2 \frac{1}{2} \bar \nabla_{i_5} \bar \nabla_{i_4} \nabla_{i_3} g_{a_1 a_2} \nonumber \\
    & + [12]^2 \frac{1}{6} \left( \bar R_{i_4 i_5 i_3}^{\qquad j} \nabla_{j} g_{a_1 a_2}  + \bar R_{i_4 i_5 a_1}^{\qquad b} \nabla_{i_3} g_{b a_2} + \bar R_{i_4 i_5 a_2}^{\qquad b} \nabla_{i_3} g_{a_1 b} \right) \nonumber \\
    & + [12]^2 \frac{1}{6} \left( \bar R_{i_3 i_5 i_4}^{\qquad j} \nabla_{j} g_{a_1 a_2}  + \bar R_{i_3 i_5 a_1}^{\qquad b} \nabla_{i_4} g_{b a_2} + \bar R_{i_3 i_5 a_2}^{\qquad b} \nabla_{i_4} g_{a_1 b} \right) \nonumber \\
    & +\left\{ (\nabla_{i_5} g_{a_1 b_1}) g^{b_1 b_2} (\nabla_{i_4} g_{b_2 b_3})g^{b_3 b_4} (\nabla_{i_3} g_{b_4 a_2})\left[  - \frac{1}{24}  [12]^2\right]  + \textrm{perm}(3,4,5)\right\}\nonumber\\
    & +\left\{ (\nabla_{i_4} g_{a_1 b_1}) g^{b_1 b_2} (\nabla_{i_5} g_{b_2 b_3})g^{b_3 b_4} (\nabla_{i_3} g_{b_4 a_2})\left[ \frac{1}{8} [12]^2 \right]  + (3 \leftrightarrow 4)\right\}\nonumber\\
    & + \frac{[12]^2 \nabla^{j} g_{a_1 a_2}}{6 } \left[  (\bar R_{i_3 i_4 i_5 j} - 2 \bar R_{i_3 i_5 i_4 j} ) \right] \nonumber \\
    =& [12]^2 \frac{1}{2} \bar \nabla_{i_5} \bar \nabla_{i_4} \nabla_{i_3} g_{a_1 a_2}  = \bar \nabla_{i_5} A_{4,a_1 a_2 i_3 i_4} . 
\end{align}
Here again there are intricate cancellations between local curvature terms and curvature terms coming from factorization channels which localize in the soft limit.

As a last example for the scalar--gauge-boson theory, let us send the momenta of two scalars to zero. For the five-particle amplitude, where $p_4,p_5\rightarrow 0$, we get
\begin{align}
    \lim_{p_4,p_5 \rightarrow 0} A_{5,a_1 a_2 i_3 i_4 i_5} &= [12]^2 \frac{1}{2} \bar \nabla_{(i_4} \bar \nabla_{i_5)} \nabla_{i_3} g_{a_1 a_2} \nonumber \\
    &+ [12]^2 \frac{1}{12} \left(  \bar R_{i_3 i_4 i_5}^{\qquad j} \nabla_{j} g_{a_1 a_2} + \bar R_{i_3 i_4 a_1}^{\qquad b} \nabla_{i_5} g_{b a_2} + \bar R_{i_3 i_4 a_2}^{\qquad b} \nabla_{i_5} g_{a_1 b} \right) \nonumber \\
    &+ [12]^2 \frac{1}{12} \left(  \bar R_{i_3 i_5 i_4}^{\qquad j} \nabla_{j} g_{a_1 a_2} + \bar R_{i_3 i_5 a_1}^{\qquad b} \nabla_{i_4} g_{b a_2} + \bar R_{i_3 i_5 a_2}^{\qquad b} \nabla_{i_4} g_{a_1 b} \right) \nonumber \\
    & +\left\{ (\nabla_{i_5} g_{a_1 b_1}) g^{b_1 b_2} (\nabla_{i_4} g_{b_2 b_3})g^{b_3 b_4} (\nabla_{i_3} g_{b_4 a_2})\left[ - \frac{1}{24}  [12]^2\right]  + \textrm{perm}(3,4,5)\right\}\nonumber\\
    & +\left\{ \frac{(\nabla_{i_5} g_{a_1 b_1}) g^{b_1 b_2} (\nabla_{i_4} g_{b_2 b_3})g^{b_3 b_4} (\nabla_{i_3} g_{b_4 a_2})}{s_{15}s_{23}}\left[ \frac{1}{8} [1|p_5 p_3|2]^2 \right]  + \textrm{perm}(3,4,5)\right\}\nonumber\\
    & + \frac{[12]^2 \nabla^{j} g_{a_1 a_2}}{6 s_{345}} \left[ s_{34} (\bar R_{i_3 i_4 i_5 j} - 2 \bar R_{i_3 i_5 i_4 j} )  + s_{35} (\bar R_{i_3 i_5 i_4 j} - 2 \bar R_{i_3 i_4 i_5 j} ) \right] \nonumber \\
    \nonumber \\
    &= [12]^2 \frac{1}{2} \bar \nabla_{(i_4} \bar \nabla_{i_5)} \nabla_{i_3} g_{a_1 a_2} \nonumber \\
    & - \frac{(s_{34}-s_{35})}{2 (s_{34} + s_{35})} \bar R_{i_4 i_5 i_3}^{\qquad j}   \left[ [12]^2 \frac{1}{2} \nabla_{j} g_{a_1 a_2} \right] \nonumber \\
    & -\frac{(s_{14}-s_{15})}{2(s_{14}+s_{15})}\bar R_{i_4 i_5 a_1}^{\qquad b}\left[ [12]^2 \frac{1}{2} \nabla_{i_3} g_{b a_2}\right]  \nonumber\\
    & -\frac{(s_{24}-s_{25})}{2(s_{24}+s_{25})}\bar R_{i_4 i_5 a_2}^{\qquad b}\left[ [12]^2 \frac{1}{2} \nabla_{i_3} g_{a_1 b}\right] \nonumber \\
    &= \bar \nabla_{(i_4} \bar \nabla_{i_5)} A_{3,a_1 a_2 i_3} + \frac{1}{2}\sum_{a=1}^{3} \frac{p_{a}\cdot (p_{4}-p_{5})}{p_{a}\cdot (p_{4} + p_{5})} \bar R_{i_4 i_5 \;\;\;\;\alpha_a}^{\quad\; \beta_a}\; A_{3,\cdots \beta_a \cdots}
    . 
\end{align}
This is the double soft theorem in \cref{eq:double_scalar_soft_theorem} where the scalars interact with gauge bosons.

\subsection{Scalars, fermions, and gauge bosons}\label{sec:dipole-example}

In this example, we will combine scalars, fermions, and gauge bosons in the same scattering amplitude. The relevant Lagrangian is 
\begin{align}
    \Lagr =& \frac{1}{2} h_{IJ}(\phi) (\partial_{\mu} \phi)^{I} (\partial^{\mu} \phi)^{J}
    + i \frac{1}{2} k_{\bar p r}(\phi) (\bar \psi^{\bar p}\overset{\leftrightarrow}{\slashed \partial} \psi^{r})
    + i \omega_{\bar p r I}(\phi) (\bar \psi^{\bar p} \gamma_{\mu} \psi^{r}) (\partial^{\mu} \phi)^{I} \nonumber \\
    & - \frac{1}{4} g_{AB}(\phi) F_{\mu\nu}^{A} F^{B\mu\nu} + \frac{-1}{2\sqrt{2}} d_{\bar p r A}(\phi) (\bar \psi^{\bar p} \sigma^{\mu\nu} \psi^{r})F^{A}_{\mu\nu} \,,
\end{align}
where the fermions and gauge bosons couple through the dipole term $d_{\bar p r A}$. The normalization for the dipole term is chosen for later convenience. The scattering amplitudes with two negative-helicity fermions, one negative-helicity gauge boson, and zero, one, or two scalars are
\begin{align}
    \label{eq:amplitude3_dipole}
    A_{3,\bar r_1 r_2 a_3} = & \langle 13 \rangle \langle 23 \rangle d_{\bar r_1 r_2 a_3} , \\
    \label{eq:amplitude4_dipole}
    A_{4,\bar r_1 r_2 a_3 i_4} = & \langle 13 \rangle \langle 23 \rangle \bar\nabla_{i_4} d_{\bar r_1 r_2 a_3}, \\
    A_{5,\bar r_1 r_2 a_3 i_4 i_5}= &  \langle 13 \rangle \langle 23 \rangle \bar\nabla_{(i_5} \bar\nabla_{i_4)} d_{\bar r_1 r_2 a_3} \nonumber \\
    &+\left(\langle 13 \rangle \langle 23\rangle \frac{p_1 \cdot (p_4-p_5)}{2p_2\cdot p_3}+\frac{\langle1 | p_{4}p_{5}-p_{5}p_{4}|3\rangle \langle 23\rangle }{4p_2\cdot p_3}\right)\bar R_{i_4 i_5 \bar r_1 }^{\quad \,\,\,\,\,\,\bar s} d_{\bar s r_2 a_3} \nonumber \\
    &+\left(\langle 13 \rangle \langle 23\rangle \frac{p_2 \cdot (p_4-p_5)}{2p_1\cdot p_3}+\frac{\langle2 | p_{4}p_{5}-p_{5}p_{4}|3\rangle \langle 13\rangle }{4p_1\cdot p_3}\right)\bar R_{i_4 i_5 r_2}^{\quad \quad s} d_{\bar r_1 s a_3} \nonumber \\
    & +\bigg\{  d_{\bar r_1 r_2 b} g^{bc} (\nabla_{i_5} g_{cd}) g^{de} (\nabla_{i_4} g_{ea_3}) \times  \bigg[
    \langle 23 \rangle \langle 13 \rangle \frac{s_{34} - s_{35} - s_{45}}{8 s_{345}} 
    \nonumber \\ & 
    \qquad\qquad\qquad\qquad + \left( \langle 25 \rangle \langle 31 \rangle + \langle 23 \rangle \langle 51 \rangle \right) \frac{s_{345} + s_{34} - s_{35}}{8 s_{345} s_{34}}  \langle 34 \rangle [54] 
    \nonumber \\ & 
    \qquad\qquad\qquad\qquad - \left( \langle 24 \rangle \langle 51 \rangle + \langle 25 \rangle \langle 41 \rangle \right) \frac{\langle 34 \rangle \langle 35 \rangle [45]^2}{8 s_{345} s_{34}} \bigg] + (4\leftrightarrow 5) \bigg\}.
\end{align}
Here we see the interplay between the various sectors in the full field-space geometry. Consider the four-particle amplitude. It depends on the covariant derivative of the dipole coupling, which is
\begin{align}
    \bar\nabla_{i} d_{\bar p r a}  = \partial_{i} d_{\bar p r a} - \bar \Gamma_{i \bar p}^{\bar s} d_{\bar s r a} - \bar \Gamma_{i r}^{s} d_{\bar p s a}- \bar \Gamma_{i a}^{b}d_{\bar p r b}.  
\end{align}
Recall from \cref{eq:fermion_Christoffel1,eq:fermion_Christoffel2,eq:gauge_Christoffel} that
\begin{align}
    \bar \Gamma^{p}_{Ir} &= k^{p\bar s} \omega^{+}_{\bar s r I},  \\
    \bar \Gamma^{\bar p}_{I\bar r} &= - \omega^{-}_{\bar r s I} k^{s \bar p}, \\
    \bar \Gamma^{a}_{ib} &= \frac{1}{2} g^{ac} (\nabla_{i} g_{cb}) \,.
\end{align}

Now we can investigate the soft limits of these amplitudes. Taking the soft limit $p_4\rightarrow 0$ in the four-particle amplitude in \cref{eq:amplitude4_dipole}, we immediately land on the covariant derivative of the three-particle amplitude in \cref{eq:amplitude3_dipole}. 

The soft limit $p_5 \rightarrow 0$ of the five-particle amplitude is a bit more involved. By collecting all the terms, we find that 
\begin{align}
    \lim_{p_5\rightarrow 0} A_{5,\bar r_1 r_2 a_3 i_4 i_5}= & \langle 13 \rangle \langle 23 \rangle  \left( \bar\nabla_{(i_5} \bar\nabla_{i_4)} d_{\bar r_1 r_2 a_3}   \right) \nonumber \\
     &+ \langle 13 \rangle \langle 23 \rangle  \left( 
     \frac{1}{2} \bar R_{i_4 i_5 \bar r_1 }^{\quad \,\,\,\,\,\,\bar s} d_{\bar s r_2 a_3} + \frac{1}{2} \bar R_{i_4 i_5 r_2}^{\quad \quad s} d_{\bar r_1 s a_3} - \frac{1}{2} \bar R_{i_4 i_5 a_3}^{\quad \quad b} d_{\bar r_1 r_2 b} \right) \nonumber \\
    =& \langle 13 \rangle \langle 23 \rangle \bar\nabla_{i_5} \bar\nabla_{i_4} d_{\bar r_1 r_2 a_3} = \bar\nabla_{i_5} A_{4,\bar r_1 r_2 a_3 i_4} \,. 
\end{align}
This is the geometric soft theorem. 

Finally, we look at the double soft limit of the five-particle amplitude. The scalar double soft limit is
\begin{align}
    \lim_{p_4,p_5\rightarrow 0} A_{5,\bar r_1 r_2 a_3 i_4 i_5}= & \langle 13 \rangle \langle 23 \rangle \bar\nabla_{(i_5} \bar\nabla_{i_4)} d_{\bar r_1 r_2 a_3} \nonumber \\
    &+\left(\langle 13 \rangle \langle 23\rangle \frac{p_1 \cdot (p_4-p_5)}{2p_1\cdot (p_4 +p_5)}\right)\bar R_{i_4 i_5 \bar r_1 }^{\quad \,\,\,\,\,\,\bar s} d_{\bar s r_2 a_3} \nonumber \\
    &+\left(\langle 13 \rangle \langle 23\rangle \frac{p_2 \cdot (p_4-p_5)}{2p_2\cdot (p_4+p_5)}\right)\bar R_{i_4 i_5 r_2}^{\quad \quad s} d_{\bar r_1 s a_3} \nonumber \\
    &-\left(\langle 13 \rangle \langle 23\rangle \frac{p_3 \cdot (p_4-p_5)}{2p_3\cdot (p_4+p_5)}\right)\bar R_{i_4 i_5 a_3}^{\quad \quad b} d_{\bar r_1 r_2 b} \nonumber \\
    =& \bar \nabla_{(i_4} \bar \nabla_{i_5)} A_{3,a_1 a_2 i_3} + \frac{1}{2}\sum_{a=1}^{3} \frac{p_{a}\cdot (p_{4}-p_{5})}{p_{a}\cdot (p_{4} + p_{5})} \bar R_{i_4 i_5 \;\;\;\;\alpha_a}^{\quad\; \beta_a}\; A_{3,\cdots \beta_a \cdots}
    . 
\end{align}
This agrees with \cref{eq:double_scalar_soft_theorem}.

\subsection{Massive gauge bosons}
\label{sec:massive-gauge-example}
Lastly, we will consider an example with massive gauge bosons. To keep the expressions manageable, we restrict to a flat field-space geometry for the gauge fields with the Lagrangian
\begin{align}
    \Lagr =& \frac{1}{2} h_{IJ}(\phi)(D_{\mu} \phi)^{I} (D^{\mu} \phi)^{J} - V(\phi)
    - \frac{1}{4} \delta_{AB} F_{\mu\nu}^{A} F^{B\mu\nu} .
\end{align}
Furthermore, we assume the following spectrum: scalars with arbitrary masses $m_j$ (could be either massive or massless) and massive gauge bosons with mass $m$. The three- and four-point amplitudes for massive gauge bosons and massless scalars are
\begin{align}
    \label{eq:A3_massive}
    A_{3, a_1 a_2 i_3}& =\frac{\nabla_{i_3}m^2_{a_1a_2}}{m^2}\langle \bS 1 \bS2\rangle[\bS2 \bS1] \,,\\
    A_{3, a_1 i_2 i_3}& =-i\sqrt{2}\nabla_{i_3}t_{a_1i_2}\frac{\langle\bS1 |p_3| \bS 1]}{m} \,, \\
    \label{eq:amplitude_massive_gauge_4pt}
    A_{4, a_1 a_2 i_3 i_4}& =\frac{1}{m^2}\nabla_{i_4}\nabla_{i_3}m^2_{a_1 a_2}\langle \bS 1 \bS2\rangle[\bS2 \bS1] +\sum_j\frac{\nabla_{i_4}{V_{i_3}}^j}{s_{34}-m_{j}^2}\frac{\nabla_{j}m^2_{a_1 a_2}}{m^2}\langle \bS 1 \bS2\rangle[\bS2 \bS1]\nonumber \\
    &\quad+\frac{1}{m^2}\Bigg[\frac{1}{s_{14}-m^2}\nabla_{i_4}{m^2_{a_1}}^{b_1}\nabla_{i_3}m^2_{a_2 b_1}\Big(\langle\bS 1 \bS2\rangle[\bS2 \bS1]+\frac{1}{2m^2}\langle \bS 1|p_4|\bS 1]\langle \bS 2|p_3|\bS 2]\Big) \nonumber \\
    &\qquad \qquad +\sum_j\frac{2\nabla_{i_4}{t_{a_1}}^j\langle\bS1|p_4| \bS 1]}{s_{14}-m_{j}^2}\nabla_{i_3}t_{a_2j}\langle\bS2|p_3| \bS 2]+(1\leftrightarrow2)\Bigg] \,.
\end{align}
Note that in \cref{eq:amplitude_massive_gauge_4pt} we allow for the exchange of scalars of arbitrary masses $m_j$.

We will verify the soft theorem for the four-point amplitude  in \cref{eq:amplitude_massive_gauge_4pt}. Sending the momentum of the massless scalar $p_4$ to zero, we obtain in the soft limit
\begin{align}
\label{eq:massive_soft4}
 \lim_{p_4\rightarrow 0}A_{4, a_1 a_2 i_3 i_4}&=\frac{1}{m^2}\nabla_{i_4}\nabla_{i_3}m^2_{a_1 a_2}\langle \bS 1 \bS2\rangle[\bS2 \bS1] + \frac{\nabla_{i_4}{V_{i_3}}^j}{2 p_3 \cdot p_4}\frac{\nabla_{j}m^2_{a_1 a_2}}{m^2}\langle \bS 1 \bS2\rangle[\bS2 \bS1]\nonumber \\
    &\quad+\frac{1}{m^2}\Bigg[\frac{\nabla_{i_4}{m^2_{a_1}}^{b_1}}{2p_1 \cdot p_4}\nabla_{i_3}m^2_{a_2 b_1}\Big(\langle\bS 1 \bS2\rangle[\bS2 \bS1]+\frac{1}{2m^2}\langle \bS 1|p_4|\bS 1]\langle \bS 2|p_3|\bS 2]\Big) \nonumber \\
    &\qquad \qquad +\frac{2(\nabla_{i_4}t_{a_1I})\epsilon^{Ij}\langle\bS1|p_4| \bS 1]}{2p_1 \cdot p_4}\nabla_{i_3}t_{a_2j}\langle\bS2|p_3| \bS 2]+(1\leftrightarrow2)\Bigg].
\end{align} 
On the other hand, the soft limit is given by the soft operator in \cref{eq:soft_theorem_gauge} acting on the lower-point amplitude. Note that this requires a choice for the off-shell continuation of $A_{3, a_1 a_2 i_3}$.  As discussed in \cref{sec:gauge}, the soft theorem is independent of that particular choice, as we will see later on.

In our example, it is convenient to write the normalization of polarization vectors in terms of momenta $|p_i|=\sqrt{p_i^2}$ in \cref{eq:A3_massive}. Evaluating the soft theorem for this case, we find that
\begin{align}
\label{eq:massive_theorem4}
 \lim_{p_4\rightarrow 0}A_{4, a_1 a_2 i_3 i_4}&=\nabla_{i_4}\left(\frac{\nabla_{i_3}m^2_{a_1 a_2}}{|p_1||p_2|}\langle \bS 1 \bS2\rangle[\bS2 \bS1]\right)  + \frac{\nabla_{i_4}{V_{i_3}}^j}{2 p_3 \cdot p_4}\Big(1+p_4^\mu\frac{\partial}{\partial p_3^\mu}\Big)\left(\frac{\nabla_{j}m^2_{a_1 a_2}}{|p_1||p_2|}\langle \bS 1 \bS2\rangle[\bS2 \bS1]\right)\nonumber\\
    &\quad+\Bigg[-\sum_{\textrm{spin}}\frac{(\nabla_{i_4}m^2_{i_1A})}{2p_1 \cdot p_4}\frac{\langle\bS 1 |\epsilon^{*Ab_1}| \bS1 \rangle}{m}\Big(1+p_4^\mu\frac{\partial}{\partial p_1^\mu}\Big)\left(\nabla_{i_3}m^2_{a_2 B}\frac{\langle\bS 2 |\epsilon_{b_1}^{B}| \bS2 ]}{|p_2|}\right) \nonumber\\
    &\qquad \qquad+\frac{2i(\nabla_{i_4}t_{a_1I})\epsilon^{Ij}}{2p_1 \cdot p_4}\frac{\langle\bS1|p_4| \bS 1]}{\sqrt{2}m}\left(-i\sqrt{2}\nabla_{i_3}t_{a_2j}\frac{\langle\bS2 |p_3| \bS 2]}{|p_2|}\right)+(1\leftrightarrow2)\Bigg].
\end{align}

Comparing the two expressions, we see that first and third lines in \cref{eq:massive_soft4} match with \cref{eq:massive_theorem4}. Next, we need to implement the soft-momentum-shift operator acting on the lower-point amplitude in terms of the spinors (see ref.~\cite{Falkowski:2020aso}). One such option of a shift by soft momentum $q$ is given by
 \begin{align}
 \label{eq:softshift_spinor}
     \langle \textbf{p} | &\to \langle \textbf{p} |+\frac{\langle \textbf{p}| pq }{2m^2}  \,,  \nonumber\\
     |\textbf{p}] &\to |\textbf{p}] + \frac{q p|\textbf{p}]}{2m^2} \,.
 \end{align}
Applying the above shift with soft $p_4$ to hard momentum $p_1$, we obtain
 \begin{align}
    \Big(1+p_4^\mu\frac{\partial}{\partial p_1^\mu}\Big)\frac{\langle\bS 2 \bS 1 \rangle[\bS 1 \bS2 ]}{|p_1||p_2|} \to & \frac{\langle\bS 2 \bS 1 \rangle[\bS 1 \bS2 ]}{|p_1||p_2|}-\frac{\langle\bS 1|p_4| \bS1 ]}{2p_1^2}\frac{\langle\bS 2|p_1| \bS2 ]}{|p_1||p_2|}+\mathcal O (p_4^2) \nonumber \\
     =&\frac{\langle\bS 2 \bS 1 \rangle[\bS 1 \bS2 ]}{m^2}+\frac{\langle\bS 1|p_4| \bS1 ]}{2m^2}\frac{\langle\bS 2|p_3| \bS2 ]}{m^2}+\mathcal O (p_4^2) \,,
 \end{align}
where we used that in the soft limit, $\langle\bS2|p_3| \bS 2]=-\langle\bS2|p_1| \bS 2]+ \mathcal O (p_4)$. This matches with the second line in \cref{eq:massive_soft4}. Hence, we have verified that the soft limit of $A_{4,a_1a_2i_3i_4}$ is given by the soft theorem in the presence of massive gauge bosons.

Let us briefly comment on a different choice of an off-shell continuation of the lower-point amplitude. Suppose we chose to evaluate the soft limit by directly applying the soft operator to \cref{eq:A3_massive}, instead of using \cref{eq:massive_theorem4}. We see that the covariant derivative $\bar \nabla A_3$ will now pick up additional terms. At the same time, the soft shift \cref{eq:softshift_spinor} acting on $A_3$ will also have extra terms. Those two contributions precisely cancel, as required, and the soft theorem again agrees with \cref{eq:massive_soft4}.

All these examples demonstrate that the universal behavior of the soft limits for massless scalars is captured by the geometric soft theorem.

\section{Conclusion}
\label{sec:conclusion}

Scattering amplitudes in any effective field theory have a universal feature; they are invariant under changes of field basis. This invariance is manifest when we express all couplings in the theory as geometric structures, such as the Riemann curvature in field space. This was initially appreciated for scalars, and now this geometric picture has been extended to both fermions and gauge fields. 

The geometry also exposes new relations {\it between} scattering amplitudes. The geometric soft theorem for scalar effective field theories \cite{Cheung:2021yog} relates scattering amplitudes with different number of particles via the covariant derivative. In this paper, we complete this story by extending the geometric soft theorem to generic effective field theories with scalars, fermions, and gauge bosons. The more general soft theorem is still linked to the covariant derivative but now for the full field space.

Soft theorems in effective field theories can be leveraged to recursively calculate higher-point scattering amplitudes from lower-point amplitudes. The bad high-energy behavior of effective-field-theory amplitudes can be ameliorated via an appropriate subtraction which uses the knowledge of the soft behavior \cite{Cheung:2014dqa, Cheung:2015ota, Cheung:2016drk, Kampf:2013vha, Luo:2015tat,Elvang:2017mdq,Elvang:2018dco,Cheung:2018oki,Low:2019ynd,Kampf:2021bet,Kampf:2021tbk}. 
This also applies to general massless scalars using the geometric soft theorem in ref.~\cite{Cheung:2021yog}. Of course, in the latter case there is no free lunch. 
Information about higher-point contact terms is encoded in the Riemann curvature, which appears in the four-point amplitude when viewing the curvature as a function of the VEV. Using the more general geometric soft theorems presented here we can enroll many additional effective field theories (e.g., \cref{eq:Lagrangian_fermion_example1}) in the list of on-shell constructible theories, whose amplitudes satisfy recursion relations. We look forward to studying such recursion relations in future work.

Even though the field-space geometry has proven valuable for understanding effective field theories, there is still a larger landscape of invariances which is not accounted for. Namely, field redefinitions with derivatives also leave the scattering amplitudes unchanged. A full geometric explanation of this invariance is a topic of current investigations \cite{Cheung:2022vnd,Cohen:2022uuw,Craig:2023wni,Craig:2023hhp,Alminawi:2023qtf,Cohen:2023ekv}. However, any extension of the geometric picture to accommodate such field redefinitions will not affect the geometric soft theorem because the derivative deformations needed to accomplish this would vanish in the soft limit. 

A natural question to ask is whether there is a version of the geometric soft theorem that holds beyond tree level. In the simpler case where singular terms in the soft limit are absent, we believe that the soft theorem remains valid at all loop orders, perhaps even non-perturbatively. In this case, the derivation is nearly identical to one derivation of the Adler zero for pions, or the geometric soft theorem for scalar effective field theories. It will be instructive to find a rigorous proof of this, and also to investigate the fate of the geometric soft theorem at loop-level when the singular terms are present.


\begin{center} 
   {\bf Acknowledgments}
\end{center}
\noindent 
We are grateful to Jonah Berean-Dutcher, Clifford Cheung, Tim Cohen, Daniel Kapec, Aneesh Manohar, Adam Tropper, and Mark Wise for useful discussions and comments on the paper. 
M.D. is supported by the DOE under grant no.~DE-SC0011632 and by the Walter Burke Institute for Theoretical Physics.

\bibliographystyle{utphys-modified}
\bibliography{bibliographySoftTheorem}

\end{document}